\documentclass[fleqn,usenatbib]{mnras}
\usepackage{newtxtext,newtxmath}

\usepackage[T1]{fontenc}
\usepackage{array}
\usepackage{booktabs}
\usepackage[table]{xcolor}
\usepackage{booktabs}
\usepackage{siunitx} 
\DeclareRobustCommand{\VAN}[3]{#2}
\let\VANthebibliography\thebibliography
\def\thebibliography{\DeclareRobustCommand{\VAN}[3]{##3}\VANthebibliography}

\usepackage{graphicx}	
\usepackage{amsmath}	

\newcommand{\teff}{\mbox{$T_{\text{eff}}$}}

\newcommand{\logg}{\mbox{$\log g$}}
\newcommand{\feh}{\mbox{[Fe/H]}}
\newcommand{\afe}{\mbox{[$\alpha$/Fe]}}

\title[Stellar Extinction and Reddening]{High-fidelity stellar extinction with \textit{Gaia} and APOGEE -- I. The method and a new extinction curve}
\author[Jie Yu et al.]{
Jie Yu$^{1,2,3,4}$\thanks{E-mail: jie.yu@nju.edu.cn (NJU)},
Luca Casagrande$^{4}$, 
John A. Taylor$^{3}$,
Ioana Ciuc\u{a}$^{5}$,
Giacomo Cordoni$^{4}$,
Ronald Drimmel$^{6}$,
\newauthor
Shourya Khanna$^{6}$,
Hiep Nguyen$^{4}$,
Tomasz R\'o\.za\'nski$^{4,7}$,
Dennis Stello$^{8,9}$,
Haibo Yuan$^{10,11}$,
Zhen Yuan$^{1,2}$
\\
\\
$^{1}$ School of Astronomy and Space Science, Nanjing University, Nanjing 210023, People's Republic of China\\
$^{2}$ Key Laboratory of Modern Astronomy and Astrophysics, Ministry of Education, Nanjing 210023, People's Republic of China\\
$^{3}$ School of Computing, Australian National University, Acton, ACT 2601, Australia\\
$^{4}$ Research School of Astronomy \& Astrophysics, Australian National University, Cotter Rd., Weston, ACT 2611, Australia\\
$^{5}$ Kavli Institute for Particle Astrophysics \& Cosmology (KIPAC), Stanford University, Stanford, CA 94305, USA\\
$^{6}$ INAF - Osservatorio Astrofisico di Torino, via Osservatorio 20, 10025 Pino Torinese (TO), Italy\\
$^{7}$ Astronomical Institute, University of Wroc\l aw, Kopernika 11, 51-622 Wroc\l aw, Poland\\
$^{8}$ School of Physics, University of New South Wales, Sydney, NSW 2052, Australia\\
$^{9}$ Sydney Institute for Astronomy, School of Physics, A28, The University of Sydney, NSW 2006, Australia\\
$^{10}$ Institute for Frontiers in Astronomy and Astrophysics, Beijing Normal University, Beijing 102206, People's Republic of China\\
$^{11}$ School of Physics and Astronomy, Beijing Normal University, No. 19 Xinjiekouwai St., Haidian District, Beijing 100875, People's Republic of China\\
}
\date{Accepted XXX. Received YYY; in original form ZZZ}

\pubyear{\the\year{}}

\begin{document}
\label{firstpage}
\pagerange{\pageref{firstpage}--\pageref{lastpage}}
\maketitle

\begin{abstract}
The scarcity of high-fidelity extinction measurements remains a bottleneck in deriving accurate stellar properties from \textit{Gaia} parallaxes. In this work, we aim to derive precise extinction estimates for APOGEE DR19 stars, establishing a new benchmark for Galactic stellar population studies. We first determine reddening by comparing observed colors—retrieved from photometric surveys or standardized synthetic magnitudes from \textit{Gaia} low-resolution blue and red photometer (BP/RP) spectra—to intrinsic colors predicted via an \texttt{XGBoost} model.  The model is trained on  minimally reddened stars to  infer intrinsic colors and their associated uncertainties, using APOGEE stellar parameters ($T_{\rm eff}$, $\log g$, [Fe/H], and [$\alpha$/Fe]). The derived reddening values are then converted into extinctions using an anchor ratio of $A_{BP} / A_{RP} = 1.694 \pm 0.004$, derived from red-clump-like stars. Here, we provide extinction measurements in 39 filters across 10 photometric systems and introduce a new empirical extinction curve optimized for broadband passbands. Our extinction estimates ($A_V$) outperform existing results (\texttt{Bayestar19}, \texttt{StarHorse}, \texttt{SEDEX}), achieving a typical precision of $\sim$0.03~mag in $A_V$. Notably, we identify systematic deviations of up to 30\% between monochromatic and passband-integrated extinction ratios at wavelengths $\lambda > 700$~nm. This result highlights the necessity of adopting passband-specific coefficients when correcting extinction to derive stellar parameters. The derived extinction and reddening data are available to the community for download through \href {https://doi.org/10.5281/zenodo.18239586}{Zenodo}.
\end{abstract}
                           

\section{Introduction}
The release of trigonometric parallaxes for 1.5 billion stars in \textit{Gaia} DR3 \citep{lindegren2021} has fundamentally transformed stellar astrophysics, enabling stellar properties (e.g., luminosity, radius, and age) to be inferred on unprecedented Galactic scales. These quantities in turn underpin exoplanet characterization, stellar evolution modeling, and chemo-dynamical studies of the Milky Way. However, the accuracy of these inferred parameters is fundamentally limited by interstellar extinction. Extinction, which reflects the wavelength-dependent absorption and scattering of starlight by dust grains, must be accurately accounted for before parallaxes can be reliably translated into physical properties. Extinction amplitude varies significantly with environment, sightline, and Galactic position \citep[e.g.,][]{green2019a, lallement2022, wang2025c, gontcharov2025}. Consequently, imprecise or inaccurate extinction corrections remain a key source of  uncertainty in precision astrophysics.

To mitigate this fundamental limitation, high-fidelity, star-by-star extinction measurements are urgently required across diverse dust environments, from the local neighborhood to the highly obscured Galactic bulge. Despite intensive efforts, such precise measurements remain scarce. The APOGEE DR19 release of spectroscopy provides a timely opportunity to address this gap, owing to its precise stellar parameters (\teff, $\log g$, [Fe/H], and [$\alpha$/Fe]) for nearly one million stars \citep{meszaros2025}. In this work, we prioritise APOGEE data because of its high quality and homogeneity, as incorporating spectroscopic data from other surveys introduce systematic inconsistencies \citep{yu2023}. These stellar parameters allow us to empirically establish the intrinsic colors of stars using nearby, minimally-reddened stars as a training sample. For more distant stars, their intrinsic colors can then be robustly inferred from their spectroscopic properties. The difference between the observed color and this predicted intrinsic color yields the color excess, which subsequently can be converted into extinction values assuming an extinction curve \citep{wang2019}. This technique, often referred to as the star-pair method \citep{stecher1965, massa1983, yuan2013, wang2025c}, has been widely and successfully applied for extinction and reddening determinations across many large photometric surveys \cite[e.g.,][]{yuan2013,wang2019,zhang2023}, such as SDSS \citep{alam2015}, Pan-STARRS \citep{chambers2019}, and 2MASS \citep{cutri2003}.

For the precise determination of stellar reddening and extinction, our methodology leverages the high internal precision of the APOGEE DR19 stellar parameters, which achieve typical uncertainties of 50–100 K in $T_{\rm eff}$, 0.07–0.09 dex in $\log g$, and 0.02–0.04 dex in [Fe/H] and [$\alpha$/Fe] \citep{meszaros2025}. However, APOGEE's spectroscopic footprint primarily lies in the northern hemisphere \citep{sdsscollaboration2025}, leading to substantially reduced overlap with major southern photometric surveys \cite[e.g., SkyMapper, ][]{onken2019, onken2024}. A transformative development that addresses this limitation comes from the availability of \textit{Gaia} BP/RP spectrophotometry \citep{deangeli2023}. \textit{Gaia} DR3 delivers flux- and wavelength-calibrated spectrophotometry for roughly 220 million sources, continuously spanning 330 to 1050 nm. From these spectra, synthetic magnitudes can be generated for virtually any passband within this wavelength interval \citep{gaiacollaboration2023b,deangeli2023,montegriffo2023}.
Crucially, \textit{Gaia} provides standardized synthetic photometry that is externally calibrated using standard stars, achieving the precision and internal consistency required for robust extinction measurements across diverse filter systems \citep{gaiacollaboration2023b}. By utilizing this standardized photometry, we mitigated residual systematics that may persist in the BP/RP spectra despite official corrections \citep{huang2024a}, as previously noted by the \textit{Gaia} collaboration \citep{montegriffo2023}.
Thus, even where direct broadband photometry is significantly incomplete within APOGEE's footprint, the combination of APOGEE stellar parameters and standardized \textit{Gaia} BP/RP synthetic magnitudes allows us to generate homogeneous, cross-survey magnitudes and, in turn, produce consistent extinction catalogues for many photometric systems.

As the first paper in this series, this study focuses on the derivation of high-fidelity, star-by-star extinction values for APOGEE stars and the construction of a new empirical extinction curve optimized for broadband passbands. The extinction curve describes how interstellar dust dims starlight as a function of wavelength ($\lambda$), usually expressed as the ratio $A_{\lambda}/A_{V}$, where $A_{V}$ serves as the reference extinction, though more recent parameterizations of the extinction curve have favored the monochromatic extinction at 550nm as the reference extinction.  Furthermore, the profile of the extinction curve varies along different lines of sight, driven by dust properties, such as grain size and composition \citep{green2025}. Historically, it has been assumed that extinction curves can be represented as a family of profiles, with their overall characteristics primarily determined by $R_{V} \equiv A_{V} / E(B-V)$, the ratio of total-to-selective extinction \citep{cardelli1989, fitzpatrick1999, gordon2023}. This ratio $R_{V}$ varies significantly across the Milky Way \citep{peek2013,zhangruoyi2023, green2025,Drimmel2025}. In diffuse interstellar regions it is typically $R_{V}\simeq3.1$, but it can increase to $R_{V} \approx 5$ or higher in the core region of dense molecular clouds, and can be as low as $R_{V}\simeq2.5$ toward high latitudes of the Milky Way.

Previous studies have provided canonical extinction curves across various wavelength ranges. Those include the far-ultraviolet (FUV) from $912$ to $1190\ \text{\AA}$ \citep{gordon2009}, the optical from $0.3$ to $1\ \mu\text{m}$ \citep{fitzpatrick2019}, the near-infrared (NIR) from $0.8$ to $5.5\ \mu\text{m}$ \citep{decleir2022}, and the mid-infrared (MIR) from $5$ to $32\ \mu\text{m}$ \citep{gordon2021}, as well as a comprehensive extinction curve spanning the full spectral range \citep{gordon2023}. However, because these canonical extinction curves are defined monochromatically ($A_\lambda$), simply applying them directly to a broadband filter by interpolating to the (weighted-mean) effective wavelength of the filter may introduce significant biases. This simplification ignores the intricate interplay between the full stellar spectral energy distribution (SED), the detailed extinction curve $A_\lambda$, the filter transmission profile $\Theta_\zeta(\lambda)$, and the bandpass width ($\lambda_1 - \lambda_2$). The correct extinction for a filter $\zeta$ should instead be computed as:
\begin{equation}\label{equa: ext}
A_\zeta = -2.5 \log \left( \frac{ \displaystyle \int_{\lambda_1}^{\lambda_2} 10^{-0.4\,A_\lambda}\,F_\lambda\,\Theta_\zeta(\lambda)\,d\lambda }{ \displaystyle \int_{\lambda_1}^{\lambda_2} F_\lambda\,\Theta_\zeta(\lambda)\,d\lambda } \right),
\end{equation}
which properly accounts for the shape of the stellar spectrum $F_\lambda$, the extinction curve $A_\lambda$, and the filter response over the entire passband. In this work, we aim to derive a new empirical extinction curve expressed in terms of $A_\zeta / A_V$ for a variety of photometric bands $\zeta$. By using stars that span a wide range in effective temperature, metallicity, and extinction --- and by directly measuring extinction in each band from observations --- we avoid the simplifying assumptions inherent in applying a fixed, monochromatic extinction curve onto broadband systems. 
The result is a set of empirically-grounded, band-specific reddening and extinction coefficients averaged over spectral types, which are valuable for deriving intrinsic stellar parameters from observed colours and magnitudes.

Along with this new empirical extinction curve, these precise extinction measurements open the door to several significant scientific applications.
Following this first paper in the series, our results will facilitate a revision of the asteroseismic scaling relation for the frequency of maximum power ($\nu_{\rm max}$, \citealt{brown1991, kjeldsen1995}, which is known to be affected by systematic biases \citep{hekker2020}. Since the scaling relations for $\nu_{\rm max}$ and the large frequency separation ($\Delta \nu$, \citealt{ulrich1986}) are fundamental for inferring stellar properties such as luminosity \citep{yu2016, yu2020}, they can be used in conjunction with bolometric corrections to derive extinctions and distances \citep{rodrigues2014}. Our independent extinction measurements, therefore, provide a direct benchmark to test and calibrate the $\nu_{\rm max}$ scaling relation. Beyond asteroseismology, our extinction framework can be applied to provide anchor points for the robust absolute calibration of the differential reddening of open clusters \citep{kalup2026}, enabling more reliable age determinations from isochrone fitting. Finally, by exploiting high-precision extinction measurements along many lines of sight, we can reconcile heterogeneous dust-map products \citep[e.g.,][]{drimmel2003, marshall2006, sale2014, green2019a, lallement2022} into a single, homogeneous, and continuous extinction scheme, yielding a unified catalog of the Milky Way that supports a broad range of stellar, Galactic, and interstellar-medium studies.

\section{Data}\label{sec: data}
\subsection{Stellar parameters in APOGEE DR19}\label{sec: apogee}
The Apache Point Observatory Galactic Evolution Experiment (APOGEE), part of the Sloan Digital Sky Survey (SDSS), is a large-scale spectroscopic survey designed to obtain high-resolution, near-infrared spectra of stars \citep{majewski2017}. APOGEE spectra have a resolution of $R \sim 22,500$ and cover the H-band ($1.51$--$1.69\ \mu\text{m}$), which minimizes interstellar extinction and enables the study of stars in highly obscured Galactic regions \citep{wilson2019}. As of December 2025, the second data release for the fifth phase of the SDSS (DR19) was the most comprehensive to date, providing APOGEE spectra and stellar parameters for one million stars \citep{sdsscollaboration2025}. These stars span a wide range of stellar types, including dwarfs, subgiants, and giants, and are sampled across all major Galactic components: the disk, bulge, and halo. 

From the $1,059,521$ entries in APOGEE DR19, we selected stars with available internally cross-matched $\texttt{gaia\_dr3\_source\_id}$, and then retained those with $\texttt{flag\_bad}$ unset to ensure the highest quality of stellar parameters, resulting in $757,338$ unique stars identified by their $\texttt{sdss\_id}$. For a star with multiple parameter entries due to multi-epoch observations, we calculated the weighted average values and uncertainties for $T_{\text{eff}}$, $\log g$, $[\text{Fe}/\text{H}]$, and $[\alpha/\text{Fe}]$, using weights equal to the inverse of the squared parameter uncertainties. Finally, we retained stars with $T_{\text{eff}} < 8000\ \text{K}$ (see the next section for details) and measurement completeness for all four parameters, resulting in a final sample of $746,853$ stars.

\subsection{Observed photometry from large-volume surveys}\label{sec: sed}
Using the $\texttt{gaia\_dr3\_source\_id}$ provided in APOGEE DR19, we retrieved photometry for $29$ passbands across $7$ photometric systems through the crossmatch service from the Gaia archive \citep{marrese2019}. These 7 photometric systems included $\textit{Gaia}$ DR3 ($G, BP, RP$, \citealt{riello2021}), Pan-STARRS DR1 ($g, r, i, y, z$, \citealt{chambers2019}), APASS DR9 ($B, V, g, r, i$, \citealt{henden2016}), SkyMapper DR2 ($u, v, g, r, i, z$, \citealt{onken2019}), SDSS DR13 ($u, g, r, i ,z$, \citealt{alam2015}), 2MASS ($J, H, K_S$, \citealt{cutri2003}), and ALLWISE ($W1, W2$, \citealt{cutri2021})\footnote{ALLWISE W3 and W4 bands were not considered in this work.}. To detect and remove photometric outliers, such as those caused by flares or saturation, we employed the $\texttt{SEDEX}$ pipeline for the SED fitting \citep{yu2021, yu2023}.

The SED fitting was performed in two steps. First, a blackbody distribution was fit to each observed SED under the assumption of zero extinction. This initial fitting was performed iteratively, removing one photometric outlier per iteration until no flux densities deviated from the best-fitting blackbody spectrum by more than $30\%$. Second, the SED fitting was refined using MARCS synthetic spectra \citep{gustafsson2008}, constrained by measured $T_{\text{eff}}$, $\log g$, and $[\text{Fe}/\text{H}]$ from APOGEE. Since the maximum $T_{\text{eff}}$ of the synthetic spectra is $8000\ \text{K}$, we limited our analysis to APOGEE stars with observed $T_{\text{eff}}$ below this threshold, as mentioned above in section~\ref{sec: apogee}. 

During the second step of the SED fitting process, we calculated the relative difference between the observed flux density in each bandpass and that predicted from the best-fitting model.  If the maximum relative difference exceeded 10\%, the corresponding photometric measurement was identified as an outlier and rejected. We note that a 10\% fractional flux threshold corresponds to $\sim$0.11 mag and is therefore conservative, even when modelling faint stars. This procedure—removing a single outlier per iteration—was repeated until no further outliers were identified or the remaining data points became too sparse to ensure a reliable fit. Specifically, only those SED fits with at least $5$ valid photometric measurements were retained for subsequent analysis.

Finally, to mitigate remaining photometric outliers in the infrared bands (2MASS and ALLWISE), which are susceptible to noise given their significantly lower fluxes compared to the bluer bands, we further refined the observed SEDs. This involved demanding high-quality 2MASS and ALLWISE photometry by requiring their photometric quality flags ($\texttt{Qflg}$) to be equal to $\texttt{"A"}$. After applying these final quality and $T_{\text{eff}}$ cuts, our final sample comprised $658,538$ stars, of which two thirds were red giants ($\logg<3.5$).

\subsection{Synthetic photometry based on Gaia BP/RP spectra}
Synthetic magnitudes derived from $\textit{Gaia}$ BP/RP spectra are valuable for this study due to their wide wavelength coverage and inherent homogeneity. These data are available from the $\textit{Gaia}$ archive in the table $\texttt{gaiadr3.synthetic\_photometry\_gspc}$. To ensure the reliability of these magnitudes, we applied the cuts as follows \citet{gaiacollaboration2023}. First, we filtered out stars with potential significant spectral contamination from nearby sources by restricting the corrected color excess parameter $C^\star$ to $|C^\star| < 0.05$ ($\texttt{gaiadr3.synthetic\_photometry\_gspc.c\_star} < 0.05$). Next, we selected synthetic magnitudes only when the associated flux-to-flux error ratio exceeded $30$ for the given passbands. This approach meant that, for some red giants, which dominate the APOGEE star sample, synthetic photometry was available for redder filters (higher SNRs) but not for bluer filters (lower SNRs). Finally, we restricted our analysis to synthetic photometry that had been standardized by $\textit{Gaia}$ \citep{gaiacollaboration2023}. These systems were Pan-STARRS ($g, r, i, y, z$), SDSS ($u, g, r, i, z$), Johnson-Kron-Cousins ($U, B, V, R, I$), Strömgren ($v, b, y$), and HST ($F435W, F606W, F814W$). 

In addition to these standardized sets, we also adopted SkyMapper synthetic photometry; although not formally standardized by \textit{Gaia}, we prioritized these over observed SkyMapper values for two reasons. Firstly, the \textit{Gaia} synthetic catalog provides a significantly larger sample size for our APOGEE stars compared to the available coverage in the SkyMapper Southern Survey (SMSS). More importantly, the SMSS Data Release 4 (DR4)—the latest and final release—is internally calibrated using synthetic magnitudes derived from \textit{Gaia} BP/RP spectra \citep{onken2024}. Consequently, the extinction coefficients derived in this work are directly applicable to the SkyMapper DR4 photometric system. We note that synthetic SkyMapper $u$-band photometry was excluded from this analysis; this is because while the $v, g, r, i,$ and $z$ filters are within the \textit{Gaia} BP/RP wavelength range ($\sim$330--1050 nm), the $u$ filter extends down to 300 nm, falling outside the blue-edge limit of the spectra. 

While both observed and standardised synthetic photometry were available for the SDSS and Pan-STARRS systems, we prioritized synthetic photometry for these two systems, similarly due to its availability for a larger number of stars. However, observed photometry was still utilized for SED fitting, as incorporating more passbands typically improves the quality of the fit. Consequently, our final dataset included observed photometry for four systems ($\textit{Gaia}$ DR3, APASS DR9, 2MASS, and ALLWISE) and synthetic photometry for six additional systems (Pan-STARRS, SDSS, Johnson-Kron-Cousins, SkyMapper, Strömgren, and HST), providing coverage across $39$ filters in total.

 \begin{figure}
    \includegraphics[width=\columnwidth]{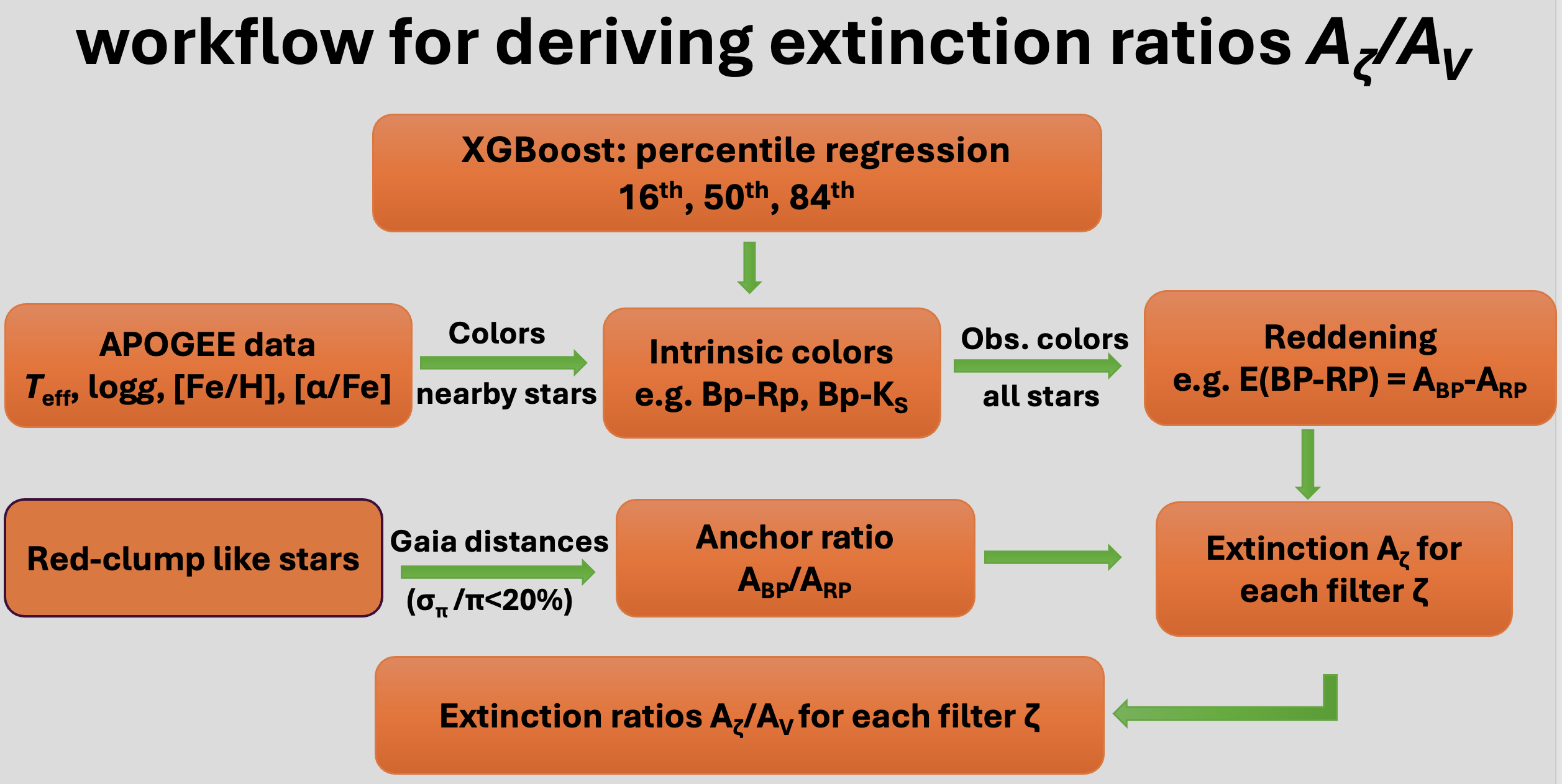}\\
    \caption{Schematic overview of the methodology used to derive extinction ratios $A_\zeta/A_V$. Intrinsic colours are inferred from APOGEE stellar parameters using XGBoost quantile regression. Reddening is measured by comparing observed and predicted intrinsic colours, while red-clump-like stars with precise Gaia distances provide an anchor extinction ratio $A_{BP}/A_{RP}$, enabling extinction estimates for individual filters and final extinction ratios.}
    \label{fig:workflow}
\end{figure}

\section{Methodology}\label{sec: method}
\subsection{Deriving Intrinsic Colors and Reddening}\label{sec: reddening}
Reddening, or color excess, is formally defined as the difference between observed and intrinsic colors. In this work, observed colors were calculated from either observed or BP/RP inferred synthetic magnitudes, as outlined in Section~\ref{sec: data}. Intrinsic colors for stars 
are not directly observable; however, they can be robustly inferred from fundamental stellar parameters ($T_{\mathrm{eff}}$, $\log g$, $[\mathrm{Fe}/\mathrm{H}]$, and $[\alpha/\mathrm{Fe}]$) \citep{chen2019}. To predict intrinsic colors (e.g., $(BP-J)_0$), we employ \texttt{XGBoost} (eXtreme Gradient Boosting), a machine-learning algorithm well suited for tackling regression tasks with tabular features \citep{chen2016}. In this model, the input features were $T_{\text{eff}}$, $\log g$, $[\text{Fe}/\text{H}]$, and $[\alpha/\text{Fe}]$, with values taken from APOGEE DR19. 
While the contribution of [$\alpha$/Fe] to broadband colors is small, it was included to account for its impact on molecular opacities and line blanketing in cool red giants, which dominate our sample. We found that a 0.5~dex increase in [$\alpha$/Fe], with all other parameters held fixed, results in an increase of  $\sim$0.03~mag in the predicted intrinsic $BP-J$ color.
The prediction labels were the intrinsic colors, approximated using stars identified as having negligible extinction. Fig.~\ref{fig:workflow} presents an overview of the methodology to derive reddening and extinction values. Below, we describe the criteria used to identify these low-extinction stars and define the training and validation sets, and then detail the \texttt{XGBoost} training procedure.

\subsubsection{Constructing Training, Validation, and Inference sets}\label{samples}
The low-extinction stellar sample used for model training and validation was constructed by selecting stars with negligible reddening values, $E(B-V) < 0.02$, based on the all-sky two-dimensional dust map of \citet{schlegel1998} (SFD) as recalibrated by \citet{schlafly2011}. To mitigate over-representation of densely populated regions in parameter space (e.g., the red clump), we partitioned the sample into bins in $T_{\mathrm{eff}}$ and [Fe/H], using bin sizes of $30\,\mathrm{K}$ and $0.2\,\mathrm{dex}$, respectively\footnote{We also tested binning in $T_{\mathrm{eff}}$ and \logg, which yielded inferior performance and was therefore not adopted.}. Each bin was capped at 300 stars, and bins containing fewer than 15 stars were discarded. A Galactic latitude cut of $|b|>10^\circ$ was applied to reduce contamination in both photometry (affecting colors) and spectroscopy (affecting stellar parameters). After applying these cuts, the final low-extinction stellar sample has a mean \texttt{SFD} reddening value of $E(B-V)=0.01$.

\begin{figure}
	\includegraphics[width=\columnwidth]{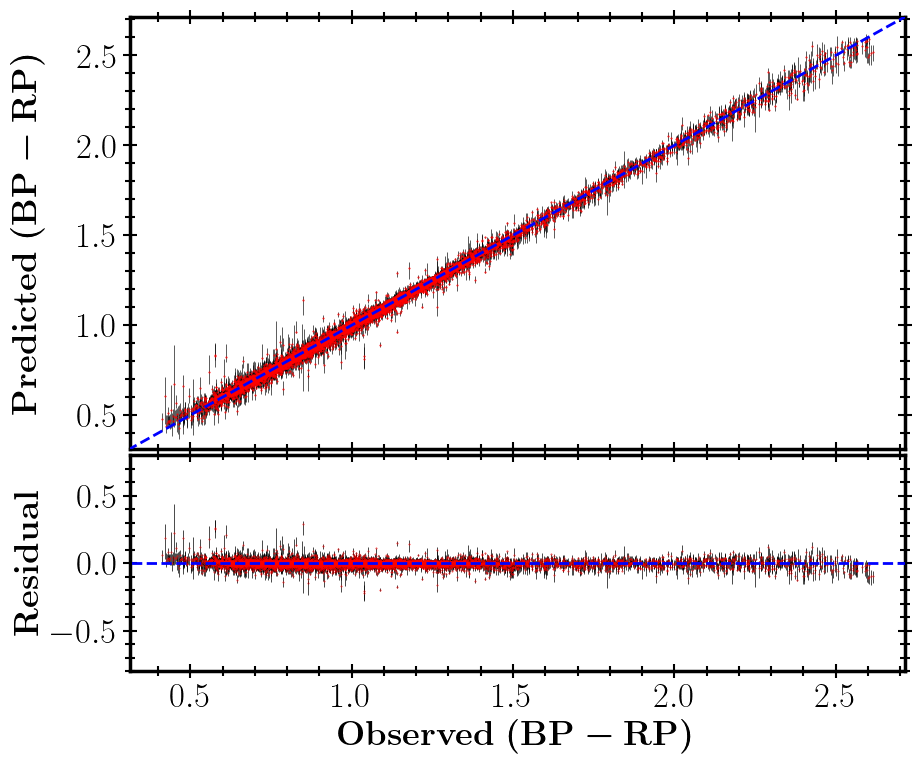}\\
    \includegraphics[width=\columnwidth]{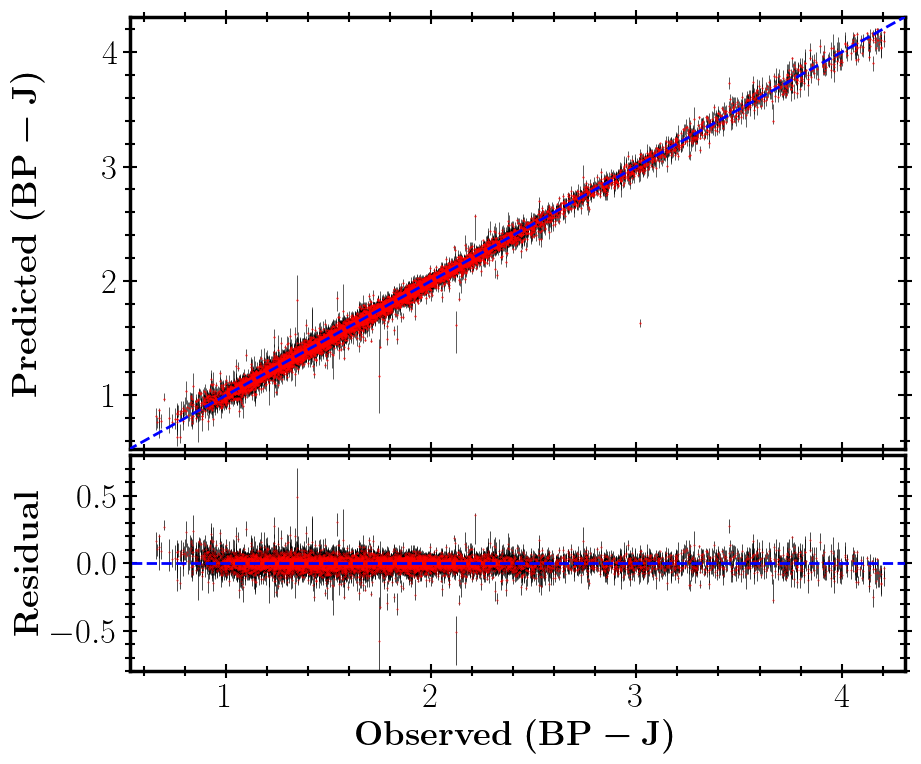}\\
    \caption{Comparison of observed and predicted intrinsic colors for 9126 stars in the validation set with negligible reddening ($E(B-V)_{\rm SFD} < 0.02$). The top panel shows $(BP - RP)$, while the bottom panel displays $(BP - J)$. Intrinsic colors are predicted using \texttt{XGBoost} regression models trained on the low-reddening stellar sample. The models utilize $T_{\rm eff}$, $\log g$, [Fe/H], and [$\alpha$/Fe] from APOGEE DR19 as input features, with the observed colors of these stars serving as target labels. $BP$ and $RP$ magnitudes are retrieved from \textit{Gaia} DR3, and $J$ magnitudes are from 2MASS. Red points represent the median predicted values, with error bars indicating the $1\sigma$ uncertainty derived from the 16th and 84th percentiles.}
    \label{fig:colorregression}
\end{figure}

\begin{figure*}
    \centering
    \includegraphics[width=0.33\textwidth]{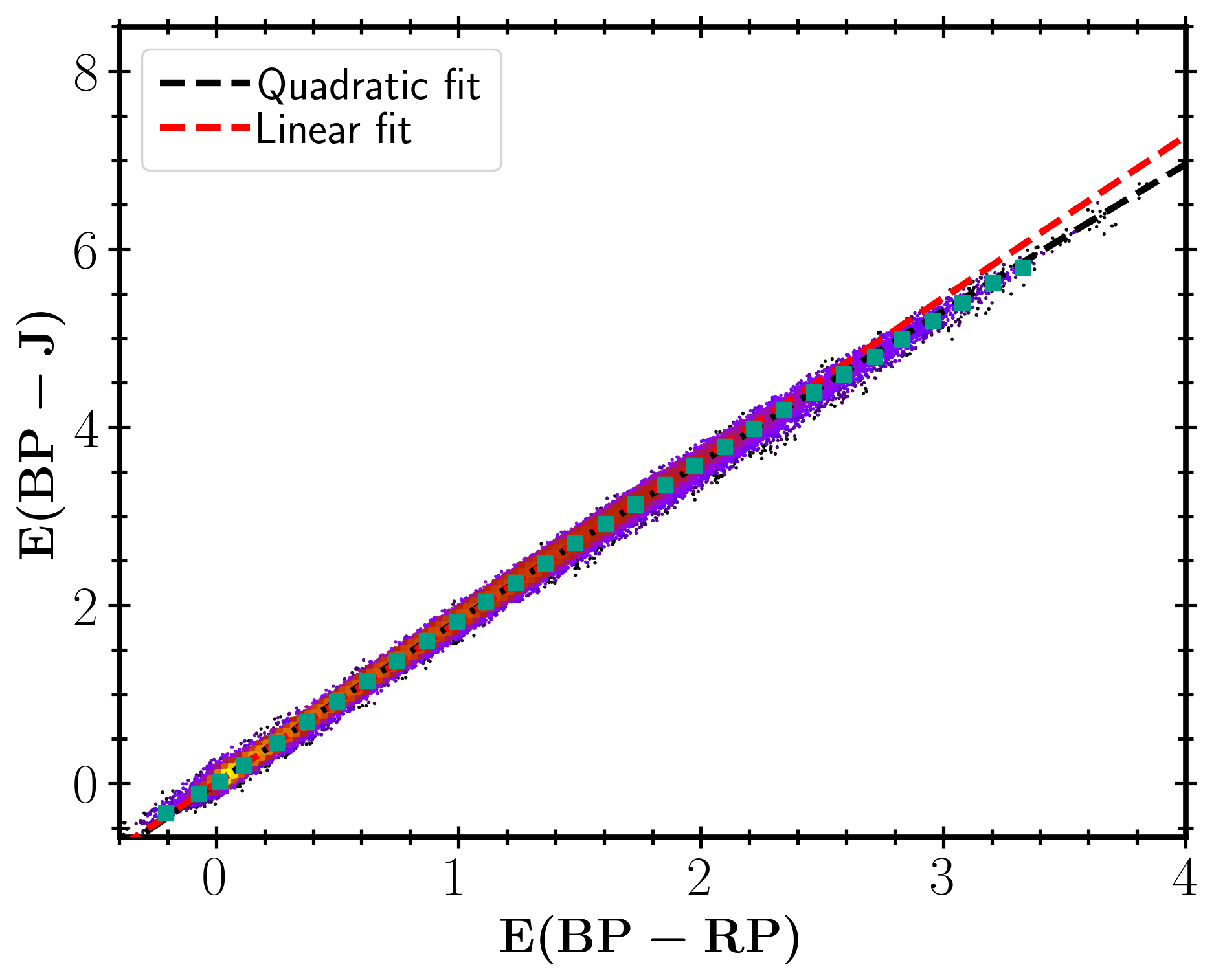}
    \hfill
    \includegraphics[width=0.33\textwidth]{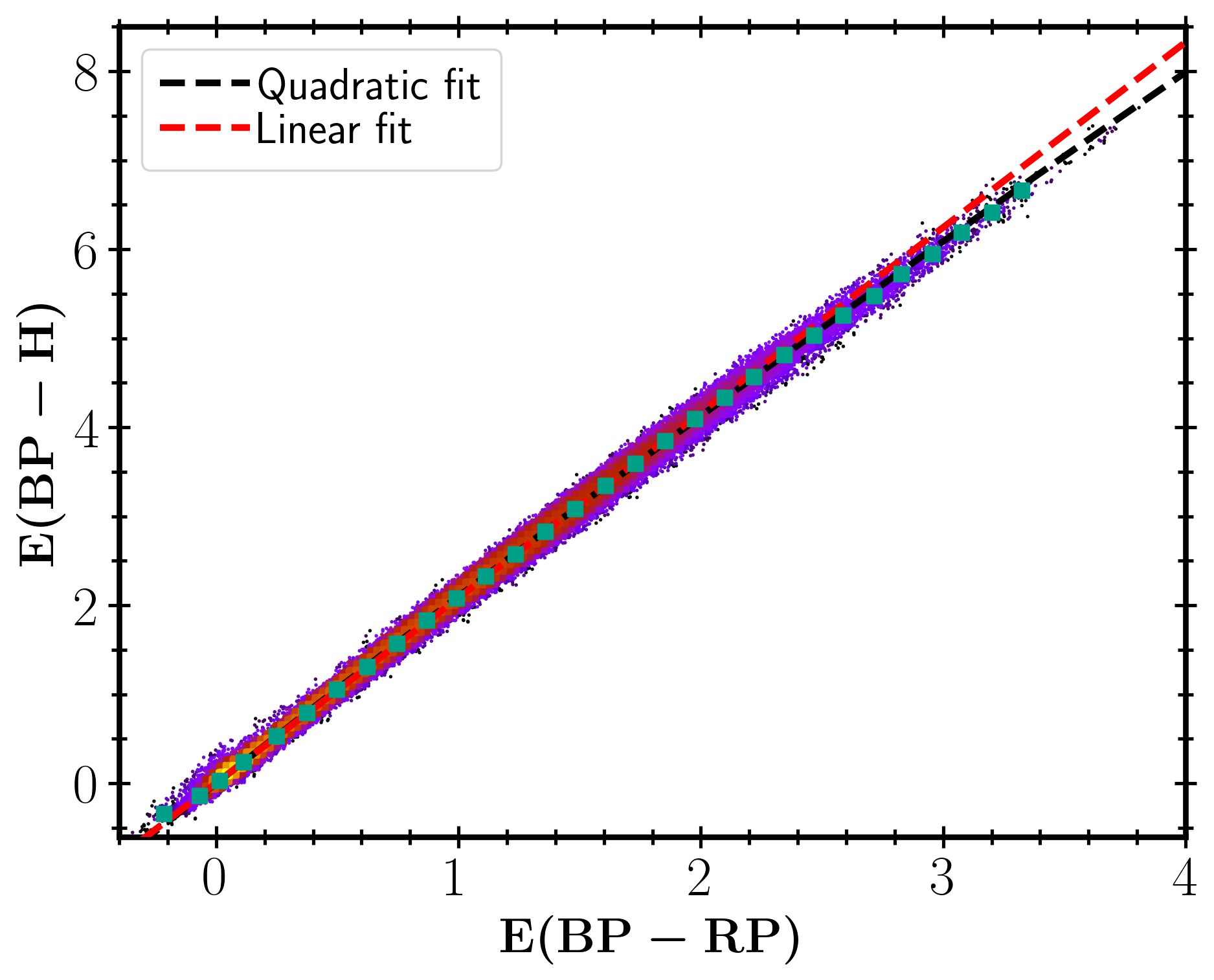}
    \hfill
    \includegraphics[width=0.33\textwidth]{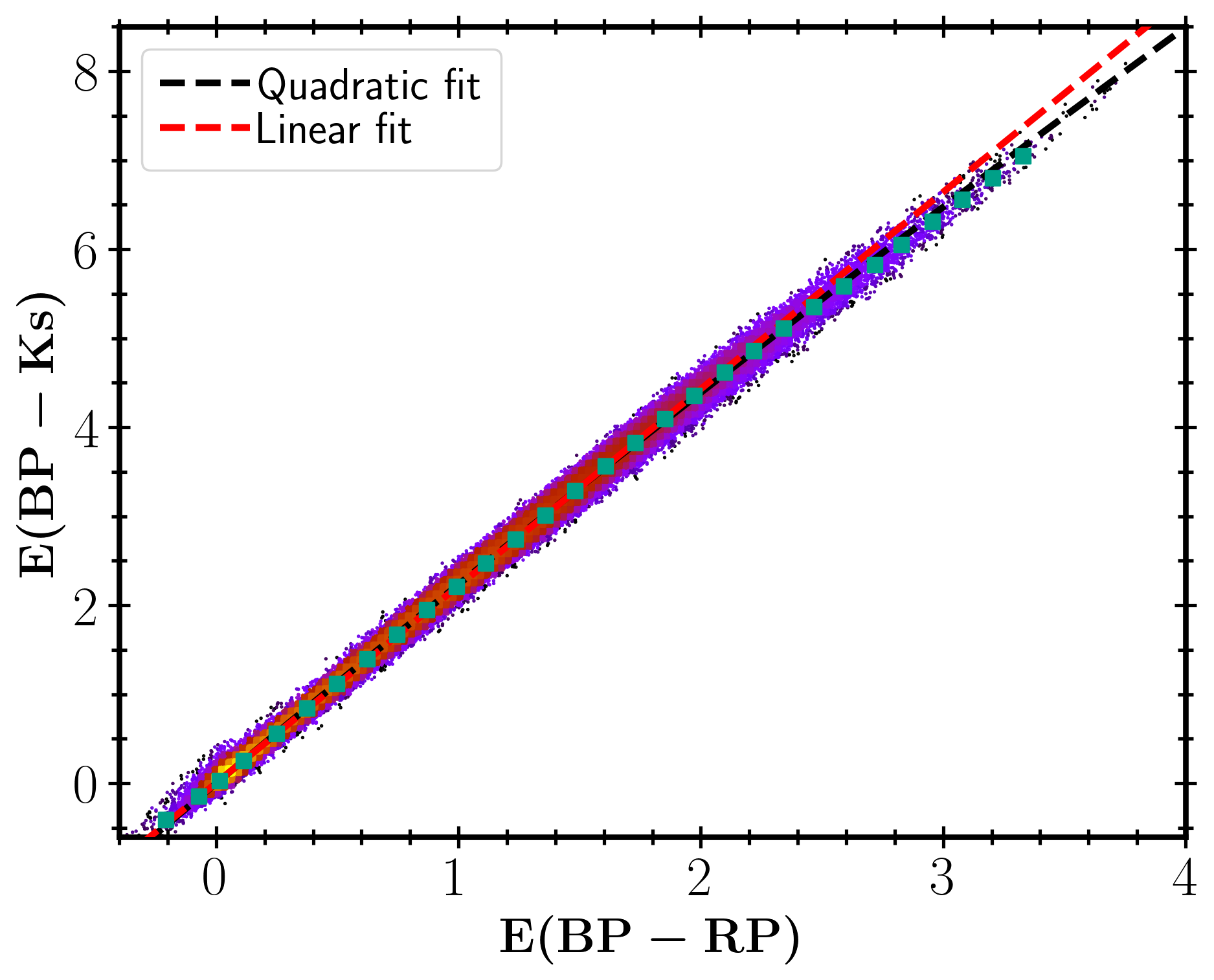}
    \includegraphics[width=0.33\textwidth]{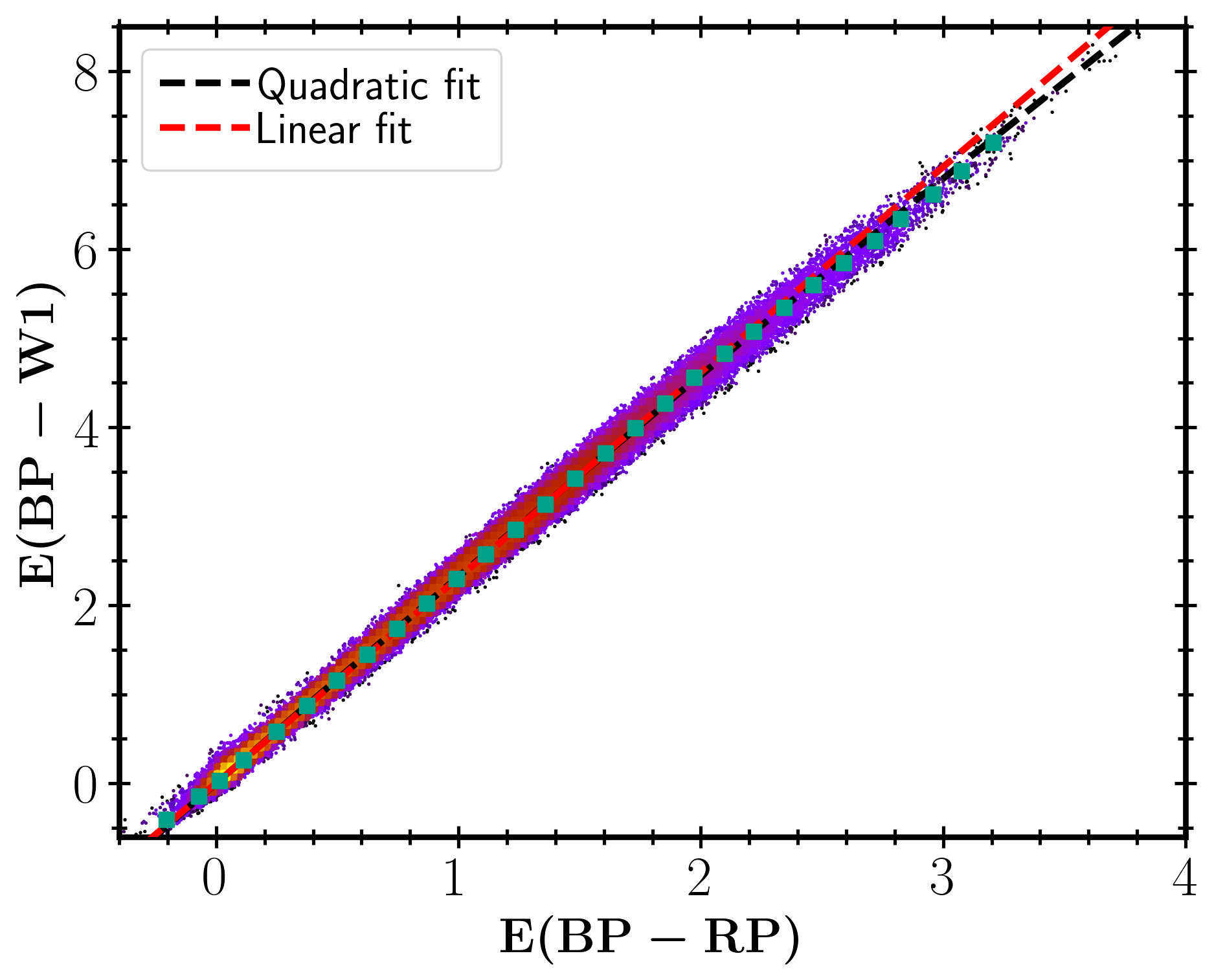}
    \includegraphics[width=0.33\textwidth]{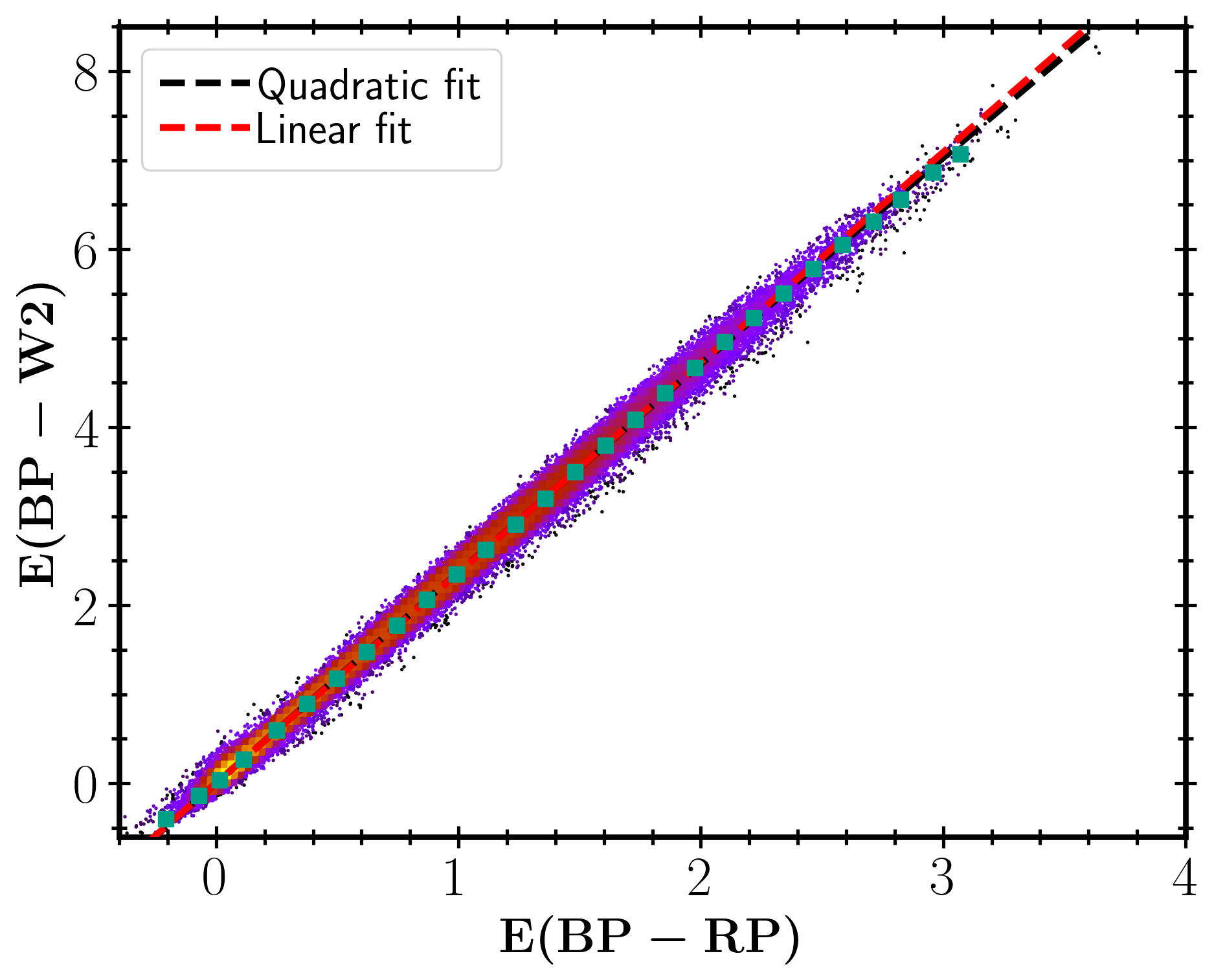}
    \caption{Comparison of reddening between \(E(BP - RP)\) and \(E(BP - \zeta)\), where \(\zeta\) corresponds to various bands. The first row shows 2MASS bands \((J, H, K_S)\), and the second row shows WISE bands \((W1, W2)\). The black dashed line represents a quadratic fit to the binned averages of the data (green squares; associated standard errors are smaller than the symbol size), while the red dashed line indicates a linear fit to the average values. The density of points is depicted using a color map, with yellow regions indicating higher densities and purple regions representing lower densities.}
    \label{fig:reddeningcomp}
\end{figure*}

The broader sample used for intrinsic color inference comprised all the APOGEE stars in the clean sample that matched the retained bin identifiers of the low-extinction sample, thereby ensuring consistency between the parameter-space coverage used for training and for inference. For model training, 90\% of the low-extinction sample was randomly assigned to the training set, while the remaining 10\% was reserved for validation. The resulting training set (82,134 stars) and validation set (9,126 stars) are sufficiently large to ensure model stability.

Finally, because our primary goal was to derive extinctions for individual filters, we restricted our reddening analysis to 38 colors rather than all possible colors. Adopting all 39 available filters would yield $39 \times 38 / 2 = 741$ colours. However, for $N$ filters, only $N-1$ colours are linearly independent, meaning that most of these 741 combinations are redundant and can be derived from a smaller fundamental set. Training separate models for all possible pairs would therefore introduce unnecessary computational overhead without adding independent information. For example, once $BP-J$ and $BP-H$ are determined, the colour $J-H$ can be directly obtained from their difference, and no additional model training is required. For this, we paired bandpasses by always using one band from $\textit{Gaia}$ $BP$ or $RP$ and the other from the remaining 38 filters, denoted $\zeta$. The choice between $BP$ and $RP$ was determined by which band yielded the largest difference in effective wavelength relative to $\zeta$ (see Table~\ref{tab:extinction} for effective wavelengths). This was because we found that training performance improved as the wavelength separation between the two bands increased.
For example, we used $u-RP$, paired from SDSS $u$ and $\textit{Gaia}$ $RP$, and $BP-J$, paired from $\textit{Gaia}$ $BP$ and 2MASS $J$.

\subsubsection{Quantile $\texttt{XGBoost}$ Implementation}\label{sec: model_training}
Extreme Gradient Boosting ($\texttt{XGBoost}$) is a robust machine learning algorithm that combines the predictions of an ensemble of decision trees constructed sequentially \citep{chen2016}. Known for its scalability, efficiency, and strong performance on low-dimensional tabular data, $\texttt{XGBoost}$ is capable of optimizing quantile loss functions and improves model generalization through regularization. Its versatility and effectiveness have been demonstrated in numerous astrophysical applications \citep[e.g.][]{andrae2023, zhang2024a}.

In this study, we employed $\texttt{XGBoost}$ for quantile regression to predict the $16^{\text{th}}$, $50^{\text{th}}$, and $84^{\text{th}}$ percentiles of intrinsic colors ($Q_{16}$, $Q_{50}$, and $Q_{84}$, respectively). These percentiles are crucial to capture the distribution of predicted colors and associated uncertainties. Separate models were trained for each of the 38 colors to ensure precision and flexibility, while maintaining a consistent architecture and hyperparameter configuration across all regressions. The median ($Q_{50}$) was adopted as the final predicted intrinsic color, and the uncertainty ($\sigma_{Q}$) was calculated as half the interpercentile range: $\sigma_{Q} = (Q_{84} - Q_{16})/2$, where $Q$ represents a color index. 

We used the official  package $\texttt{xgboost}$ \citep{chen2016}\footnote{The package implementation is available at \url{https://github.com/dmlc/xgboost}} with hyperparameters optimized for both accuracy and computational efficiency. The objective was set to $\texttt{`reg:quantileerror'}$ to model the desired percentiles, using the root mean square error as the evaluation metric on the validation set. After a thorough grid search, the optimal configuration included $5000$ boosting rounds with early stopping at $100$ rounds to prevent overfitting. The maximum tree depth was set to $4$ to balance model complexity and the capture of non-linear relationships, with a learning rate of $0.03$ for stable convergence. To enhance generalization, $80\%$ of features were randomly sampled for each tree. Fig.~\ref{fig:colorregression} shows two examples of predicted intrinsic colors for $BP-RP$ and $BP-J$, with median precisions of $0.02$ and $0.04$ mag, respectively.

Once the intrinsic colors were predicted, the reddening, or color excess $E(\zeta_1 - \zeta_2)$, was calculated using the fundamental definition:
\begin{equation}\label{eq:reddening_calc}
E(\zeta_1 - \zeta_2) = (\zeta_1 - \zeta_2)_{\text{obs}} - (\zeta_1 - \zeta_2)_{\text{intr}},
\end{equation}
where $(\zeta_1 - \zeta_2)_{\text{obs}}$ and $(\zeta_1 - \zeta_2)_{\text{intr}}$ are the observed and predicted intrinsic colors, respectively. The uncertainty in $E(\zeta_1 - \zeta_2)$ was estimated through standard error propagation, considering the uncertainties in both the observed colors (from catalog errors) and the predicted colors ($\sigma_{C}$).

\begin{figure}
	\includegraphics[width=\columnwidth]{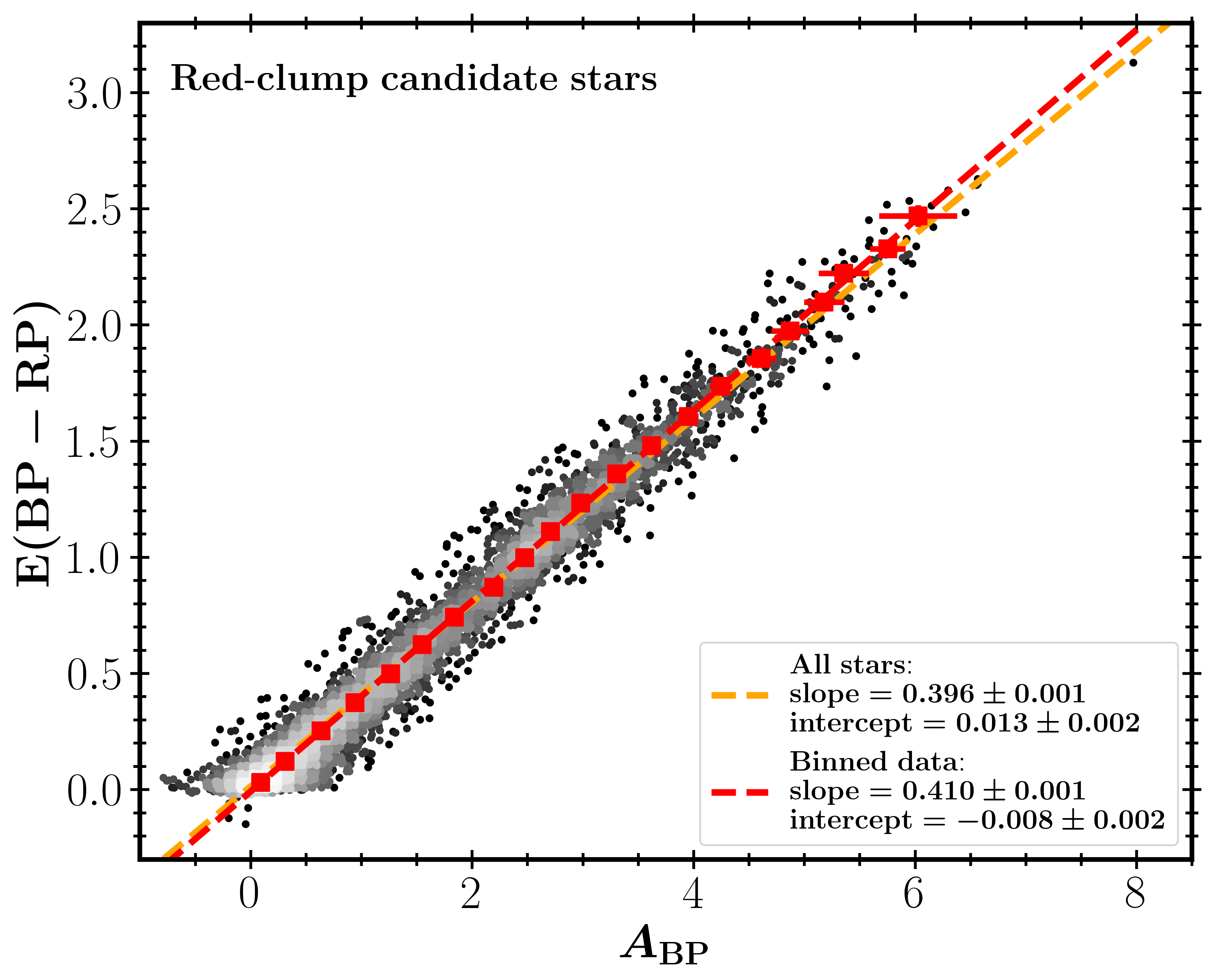}
    \caption{Extinction \(A_{BP}\) versus color excess \(E(BP - RP)\) for 8,426 red clump candidate stars (see the text for their selection), color-coded by star number density on a logarithmic scale. The orange dashed line represents the best linear fit (with $A_{BP}$ as the independent variable) to all the red clump candidate stars, while the red dashed line shows the best linear fit to the mean values binned in color excess \(E(BP - RP)\) with a bin size of 0.12. In both regression models, $A_{BP}$ was treated as the independent variable and $E(BP - RP)$ as the dependent variable. The slope and intercept of each best-fit line are annotated in the legend. The \(3\sigma\) outliers w.r.t the orange dashed line, comprising 1.7\% of the full sample, are excluded and not shown here.}
    \label{fig:anchor}
\end{figure}

Fig.~\ref{fig:reddeningcomp} illustrates the relationship between $E(BP - RP)$ and $E(BP - X)$, where $X$ corresponds to various bands (2MASS $J, H, K_s$ and WISE $W1, W2$). In addition to tight correlations, the comparison reveals a detectable non-linear relationship (curvature) at high reddening values, as indicated by the difference between linear and quadratic fits. Similar curvatures in reddening relationships have been reported in previous studies \citep[e.g.,][]{wang2019}.
This curvature arises from broadband integration effects; as shown in Equation~\ref{equa: ext}, this non-linearity is a function of both the total extinction and $T_{\rm eff}$, which primarily determines the shape of the underlying spectral energy distribution. Consequently, this effect becomes most significant in broader filters and for hotter stars at higher reddening values, where the change in the integrated flux's centroid (or effective wavelength) is most pronounced.

However, such curvatures do not clearly propagate into the extinction measurements derived from reddening values (as discussed further in Section~\ref{sec: curve}). Deriving extinction from reddening for constructing an extinction curve often involves correcting the broadband extinction $A_{\zeta}$ to extinction at the passband's effective wavelength \citep{wang2019}. However, this correction introduces a dependence on a pre-existing extinction curve, making the newly derived extinction curve not entirely independent. In this work, we avoided such corrections and instead derived the extinction for individual filters directly. This approach, as we will demonstrate in Section~\ref{sec: curve}, results in extinction ratios that do not exhibit the previously discussed curvature with $A_V$ even when $A_V$ reaches up to $\sim 8\ \text{mag}$. This means that, by design, our derived extinctions are defined for the integrated response of the full passband, rather than a single effective wavelength.

\begin{figure*}
    \includegraphics[width=0.75\textwidth]{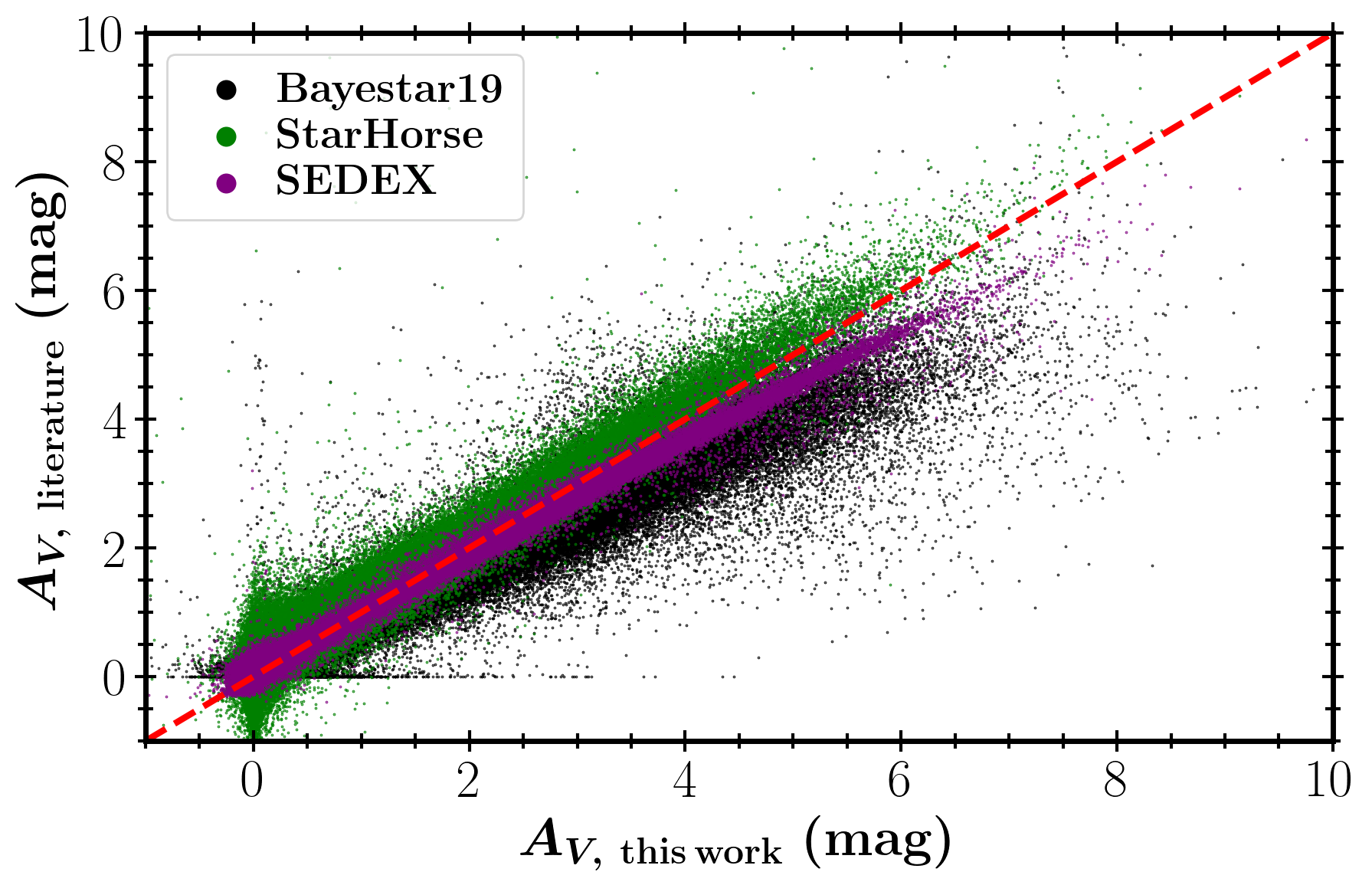}
	\includegraphics[width=0.75\textwidth]{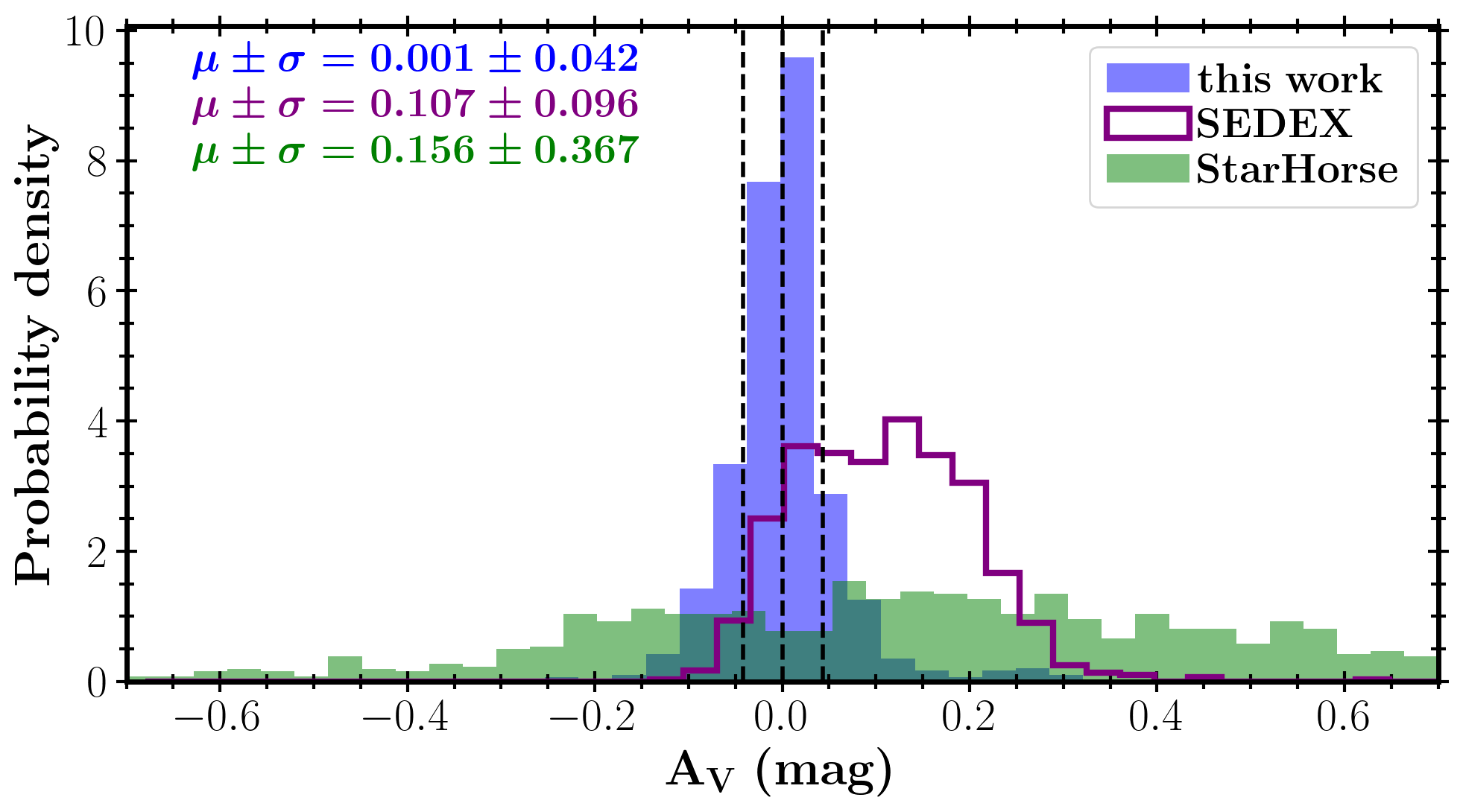}\\
    \caption{Precision and accuracy of our extinction measurements in \(A_V\). 
    \textbf{Upper panel}: Comparison of \(A_V\) values derived in this work with those from \texttt{Bayestar19} (black), \texttt{StarHorse} (green), and \texttt{SEDEX} (purple). The red dashed line represents the 1:1 relationship. \textbf{Lower panel:} Distributions of \(A_V\) derived in this work (blue) compared with those from \texttt{SEDEX} (purple) and \texttt{StarHorse} (green) for nearby stars with distances less than 70 pc. These stars are expected to experience low extinction, making the dispersion of these distributions indicative of the internal \(A_V\) precision. The mean (\(\mu\)) and 1\(\sigma\) uncertainty (calculated from the 16$^{\rm th}$ and 84$^{\rm th}$ percentiles) are shown in the plot for each method, with results for this work indicated by vertical dashed lines. The 1\(\sigma\) dispersion is 0.042 for this work, compared to \(0.096\) from \texttt{SEDEX} and \(0.367\) from \texttt{StarHorse}. Negative $A_V$ values in both panels arise due to measurement uncertainty.}
    \label{fig:avprecision}
\end{figure*}

\subsection{Deriving Extinction}\label{sec: ext}
The overall procedure for deriving extinction for all 39 passbands involved two main steps: first, establishing an accurate extinction ratio ($A_{BP}/A_{RP}$) using Red Clump (RC) like stars, and second, using this anchor point to convert all calculated reddening values into extinctions for individual filters.

 \begin{figure*}
	\includegraphics[width=0.88\columnwidth]{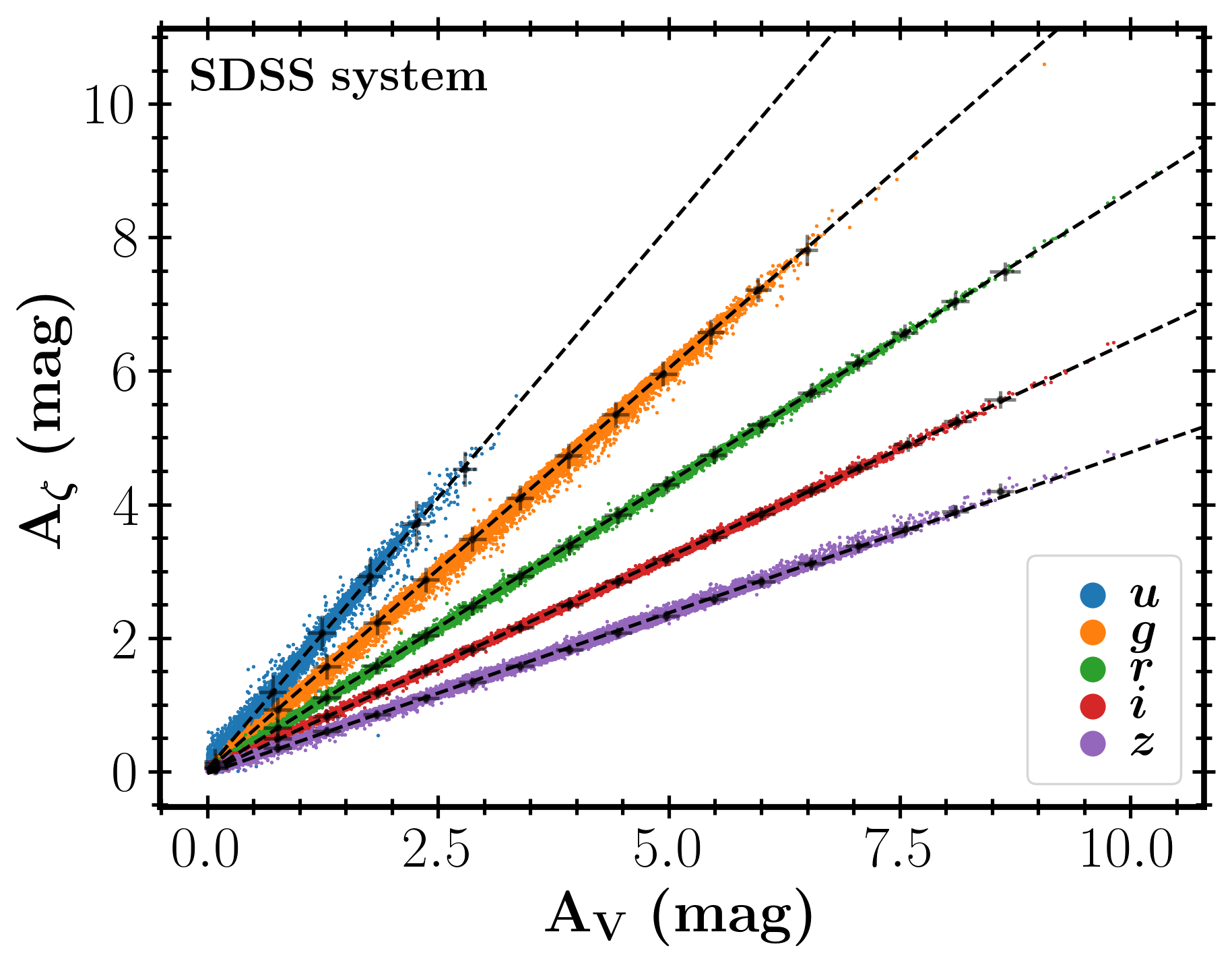}
    \includegraphics[width=0.88\columnwidth]{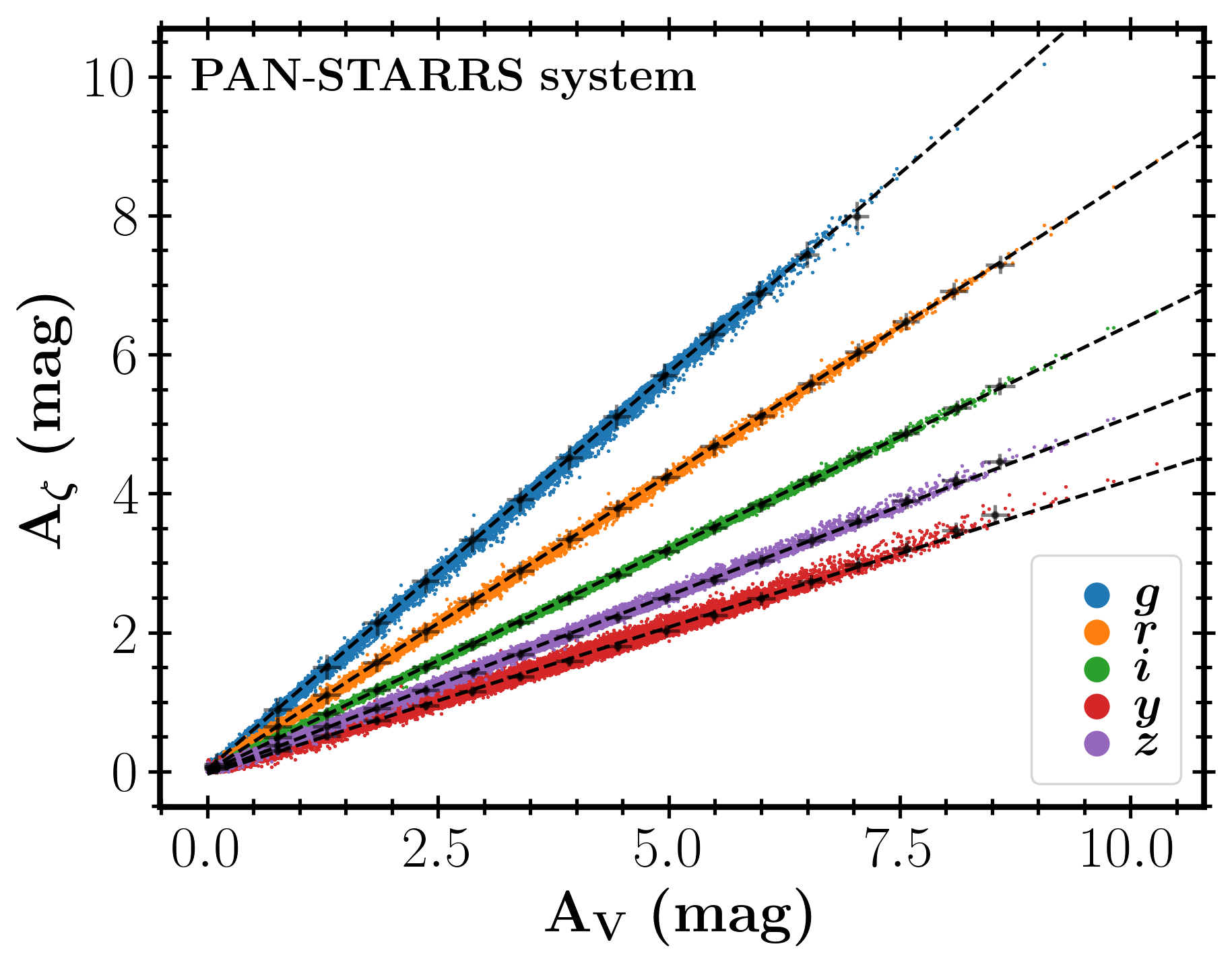}\\
    \includegraphics[width=0.88\columnwidth]{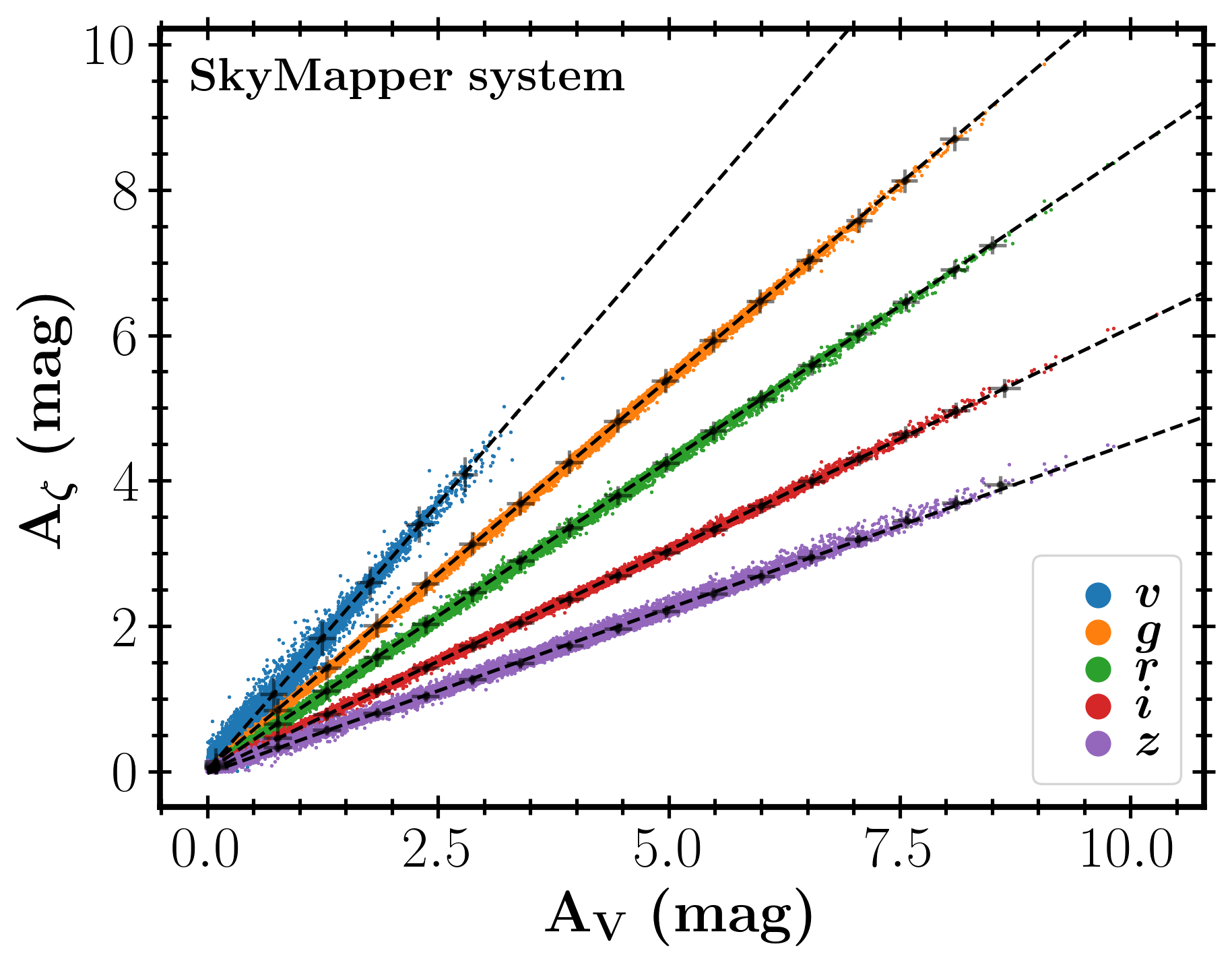}
    \includegraphics[width=0.88\columnwidth]{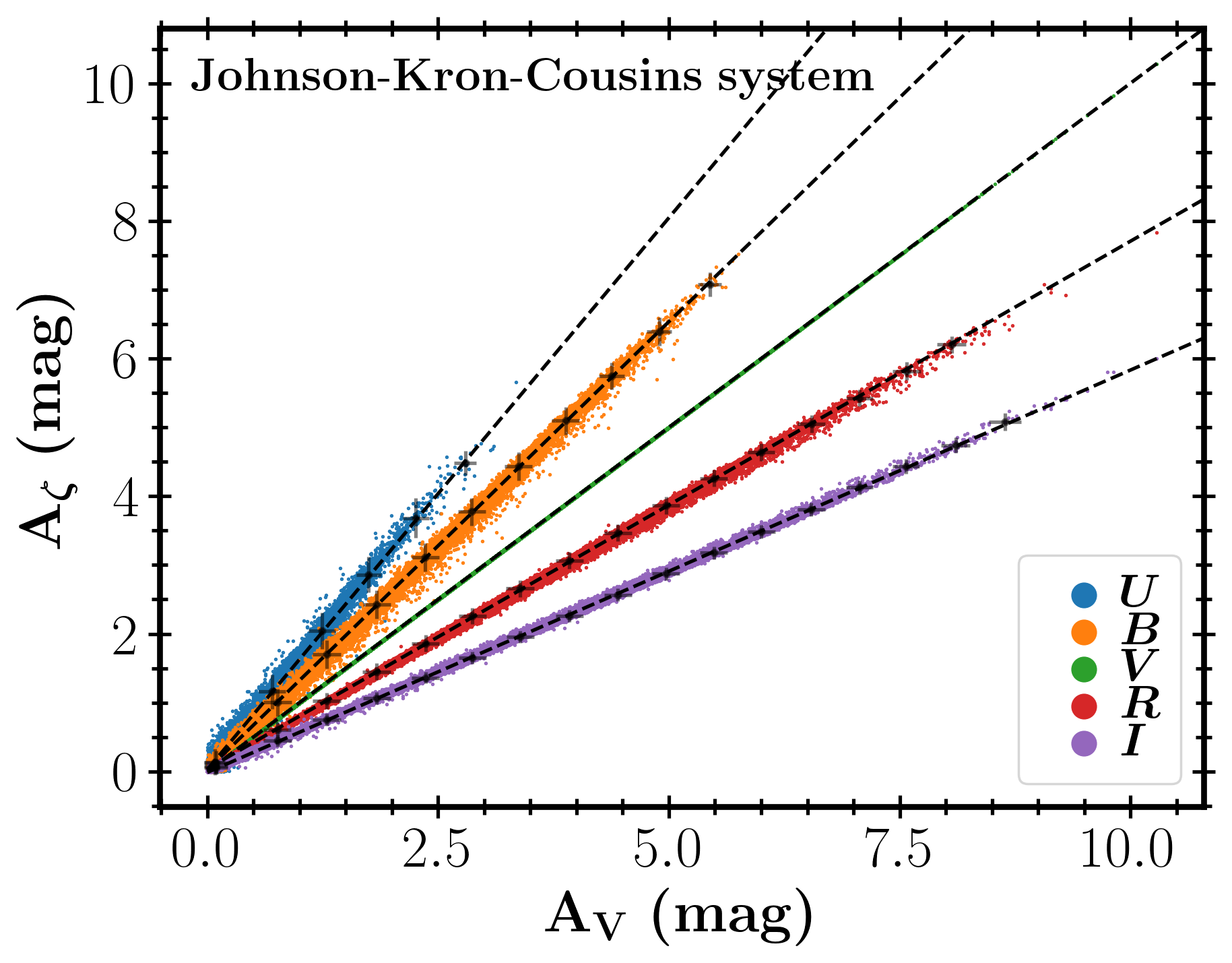}\\
    \includegraphics[width=0.88\columnwidth]{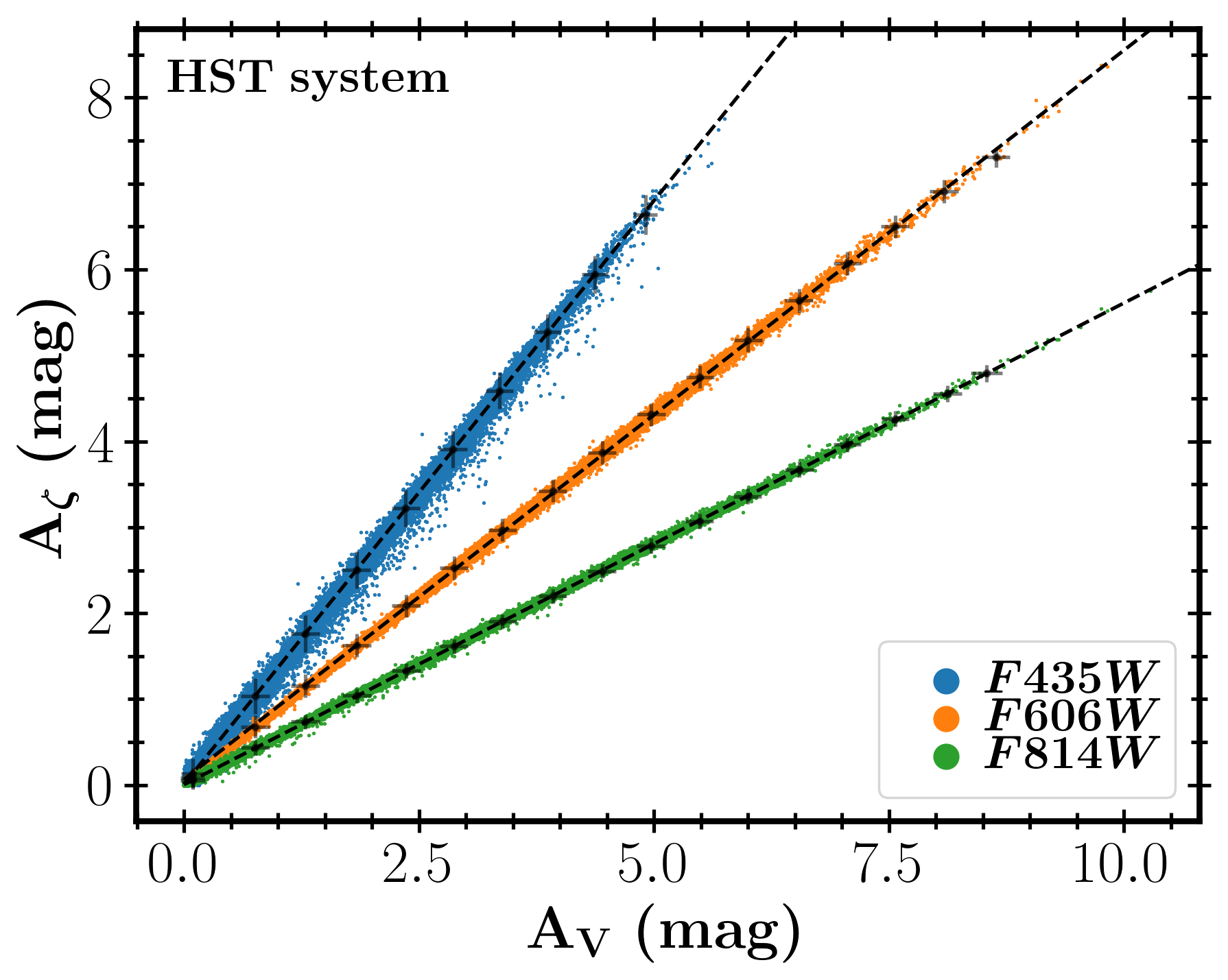}
    \includegraphics[width=0.88\columnwidth]{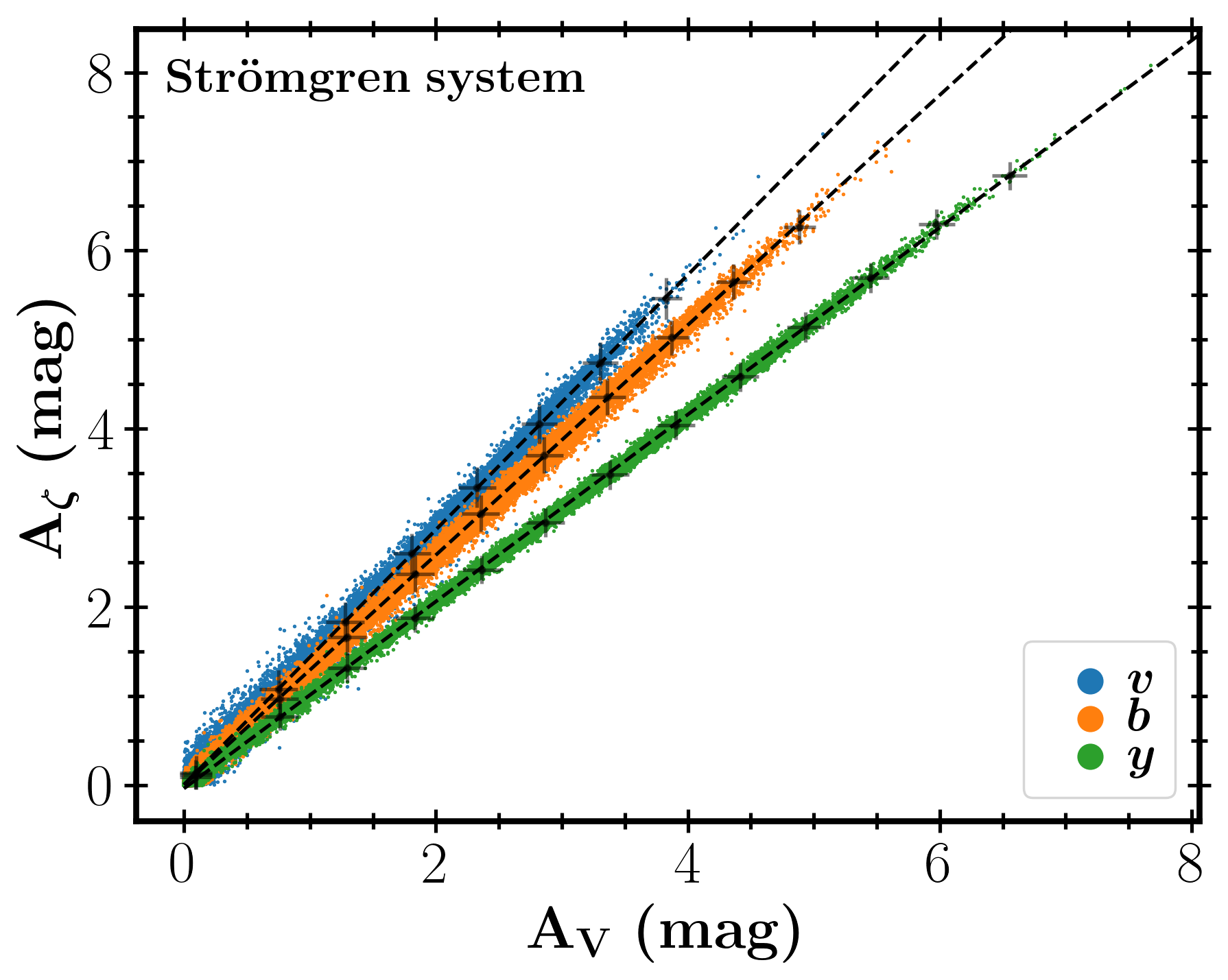}
\caption{
Extinction measurements \(A_{\zeta}\) as a function of \(A_V\) for individual passbands ($\zeta$) within six photometric systems: (a) SDSS, (b) Pan-STARRS, (c) SkyMapper, (d) Johnson-Kron-Cousins, (e) HST, and (f) Strömgren. Synthetic magnitudes derived from Gaia BP/RP spectra are used to calculate these extinctions, considering broader sky coverage compared to associated photometric surveys or observations. With the exception of SkyMapper (see the text for details), only standardized filters by \citet{gaiacollaboration2023} are included in this work, which excludes certain filters, such as HST filters different from the three shown here. Each panel displays extinction values derived for individual filters within the respective system, with different colors representing different filters. Large crosses indicate binned averages, while small dots represent individual measurements. Dashed black lines show linear fits used to derive extinction ratios \(A_{\zeta} / A_V\). The extinction measurements and their uncertainties are available for download. These data can be used for deriving extinction ratios, extinction-to-reddening ratios, and reddening ratios for user-selected passbands.}
\label{fig:extinctionratioSyn}
\end{figure*}

We determined the extinction ratio $A_{ BP}/A_{RP}$ by measuring the slope of the linear relationship between the color excess $E(BP - RP)=A_{BP}-A_{RP}$ and the extinction $A_{BP}$. Since $E(BP - RP)$ was already measured in Section~\ref{sec: reddening}, this step required the precise estimation of $A_{BP}$. For this, we selected a sample of RC-like stars, which are ideal due to their abundance and nearly uniform intrinsic absolute magnitudes \citep{khanna2025}. The selection criteria, based on APOGEE DR19 stellar parameters, were: $T_{\rm eff} = 4750 \pm 100\ \text{K}$, $\log g = 2.5 \pm 0.1\ \text{dex}$, and $[\text{Fe}/\text{H}] = 0.0 \pm 0.1\ \text{dex}$. 

We determined $A_{BP}$ through $A_{BP} = BP - M_{BP} - 5 \log_{10}(d) + 5$, 
where $BP$ is the apparent Gaia magnitude, and the remaining terms give the predicted, reddening-free apparent magnitude of a RC-like star. We derived
$M_{BP}$ from the absolute magnitude in the $K_s$ band, adopting $M_{K_{\rm S}} = -1.650\pm0.025\ \text{mag}$, a well-established value for RC stars from the literature \citep[e.g.][]{khan2023}. We then used the intrinsic color $(BP - K_{\rm s})_0$ derived from the subset of selected RC-like stars that have negligible calibrated $\texttt{SFD}$ reddening values, $E(B-V) < 0.02$ (see Section~\ref{samples}). The median intrinsic color $(BP - K_{\rm S})_0$ for this low-extinction RC sample was found to be $2.717\pm0.053\ \text{mag}$, yielding an absolute magnitude of 
$M_{BP} = (BP - K_{\rm S})_0 + M_{K_{\rm S}} =  = 1.067\pm{0.059}\ \text{mag}$. Next, we calculated $A_{BP}$ for each RC star using $M_{BP}$, observed magnitude $BP$, and distance $d$. We utilized distances from \citet{bailer-jones2021} for stars with $\textit{Gaia}$ EDR3 parallax uncertainties better than $20\%$, prioritizing photo-geometric distances when available, as they are considered more reliable than purely geometric distances \citet{bailer-jones2021}.

Fig.~\ref{fig:anchor} shows the relationship between $A_{BP}$ and $E(BP - RP)$ for the RC-like stars. To derive the extinction ratio $A_{BP}/A_{RP}$, we fit a linear model to the binned data in steps of $0.12\ \text{mag}$ (corresponding to 20 bins, each containing at least 5 stars) in $E(BP - RP)$ (red dashed line, red squares), as we preferred the binned fit to avoid bias due to the overdensity of stars with low extinction (see the grey colormap). The best fit (for the binned data) yielded a slope of $0.410 \pm 0.001$ and an intercept of $-0.008 \pm 0.002$. For comparison, the un-binned data resulted in a slope of $0.396 \pm 0.001$ and an intercept of $0.013 \pm 0.002$.
Using the slope derived from the preferred binned linear fit, we calculated the final extinction ratio to be $A_{BP}/A_{RP} = 1.694 \pm 0.004$. This value is consistent with previous estimates, such as $A_{BP}/A_{RP} = 1.686 \pm 0.016$ and $1.700 \pm 0.007$, derived in \citet{wang2019} using two different methods.

 \begin{figure*}
	\includegraphics[width=0.88\columnwidth]{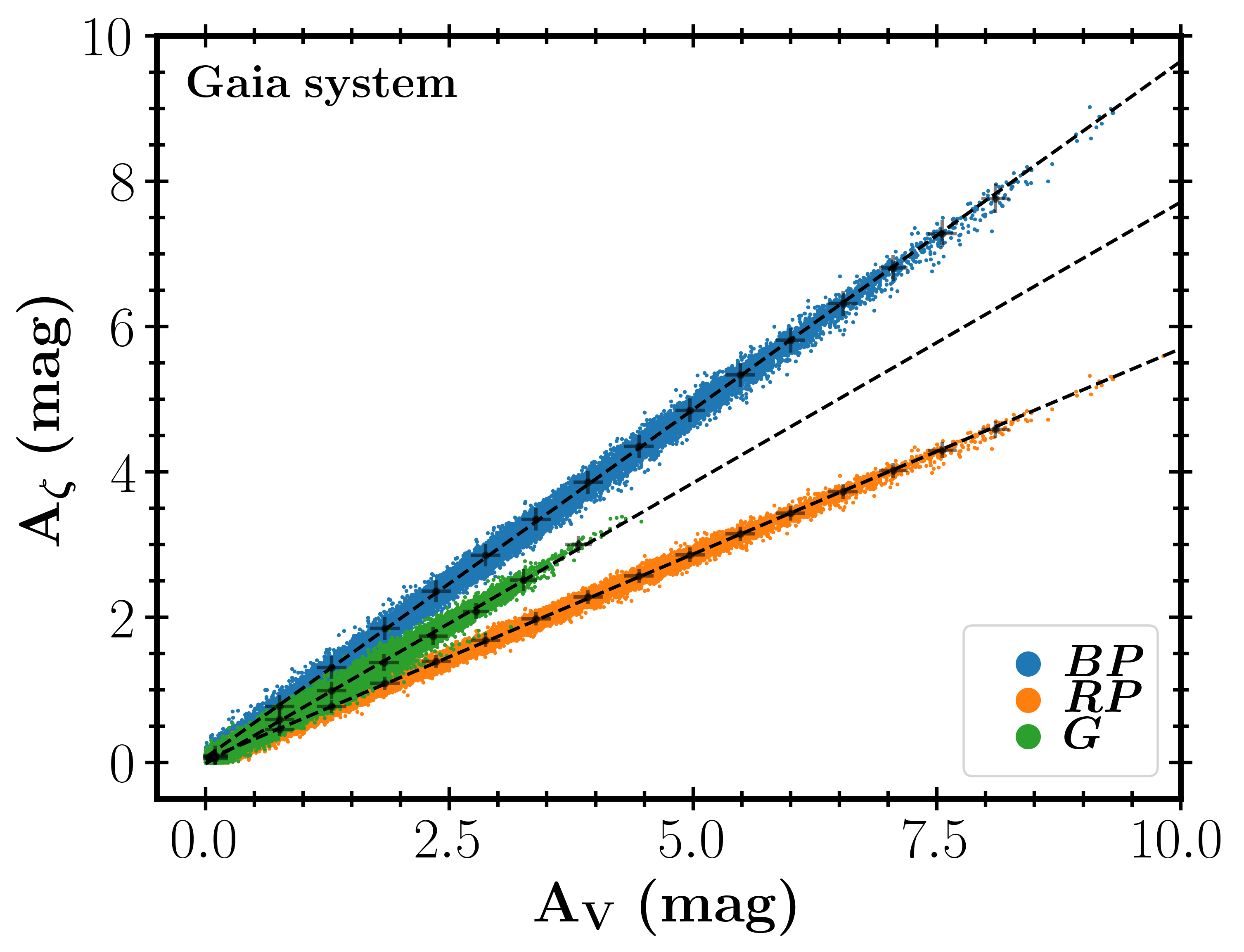}
    \includegraphics[width=0.88\columnwidth]{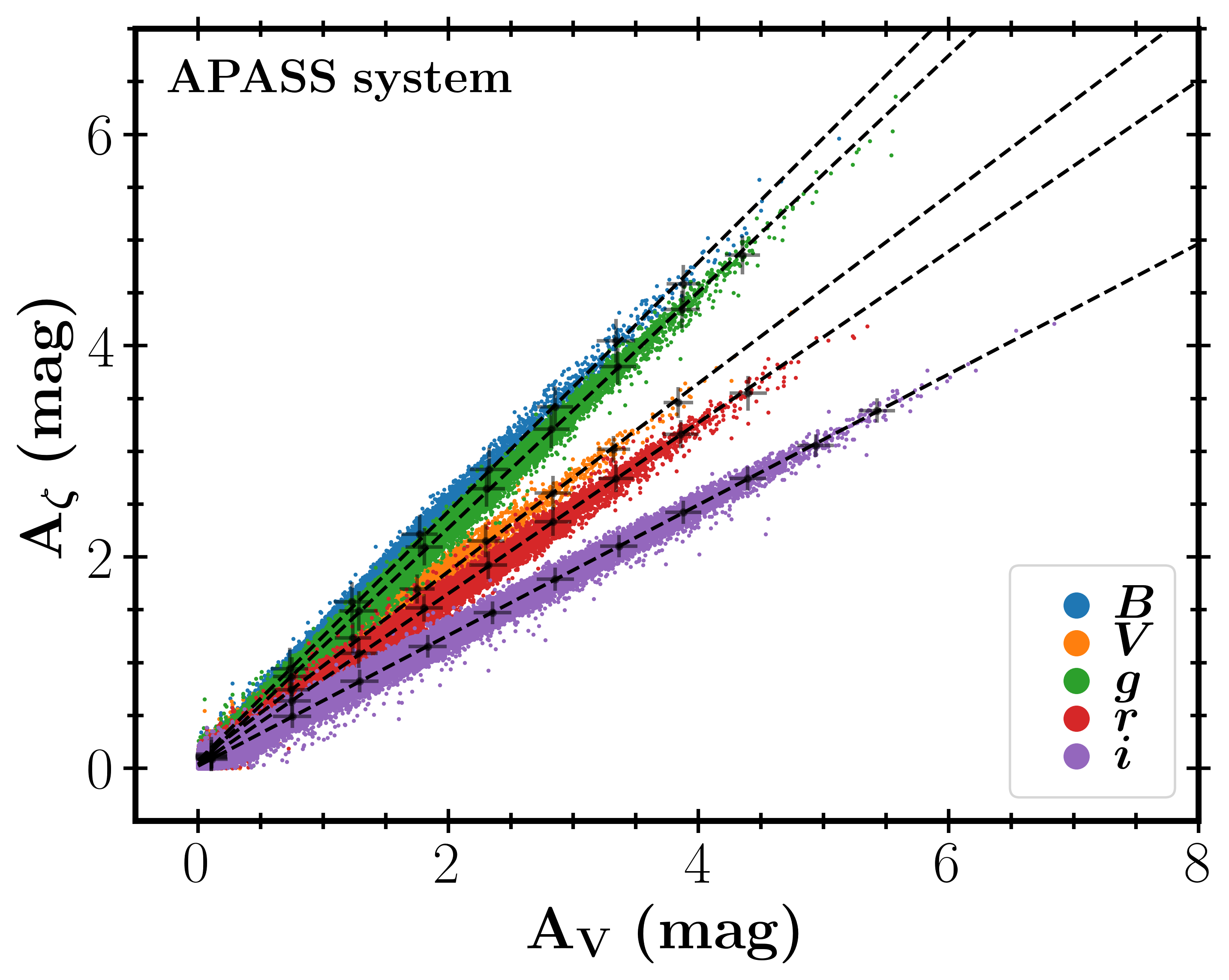}\\
    \includegraphics[width=0.88\columnwidth]{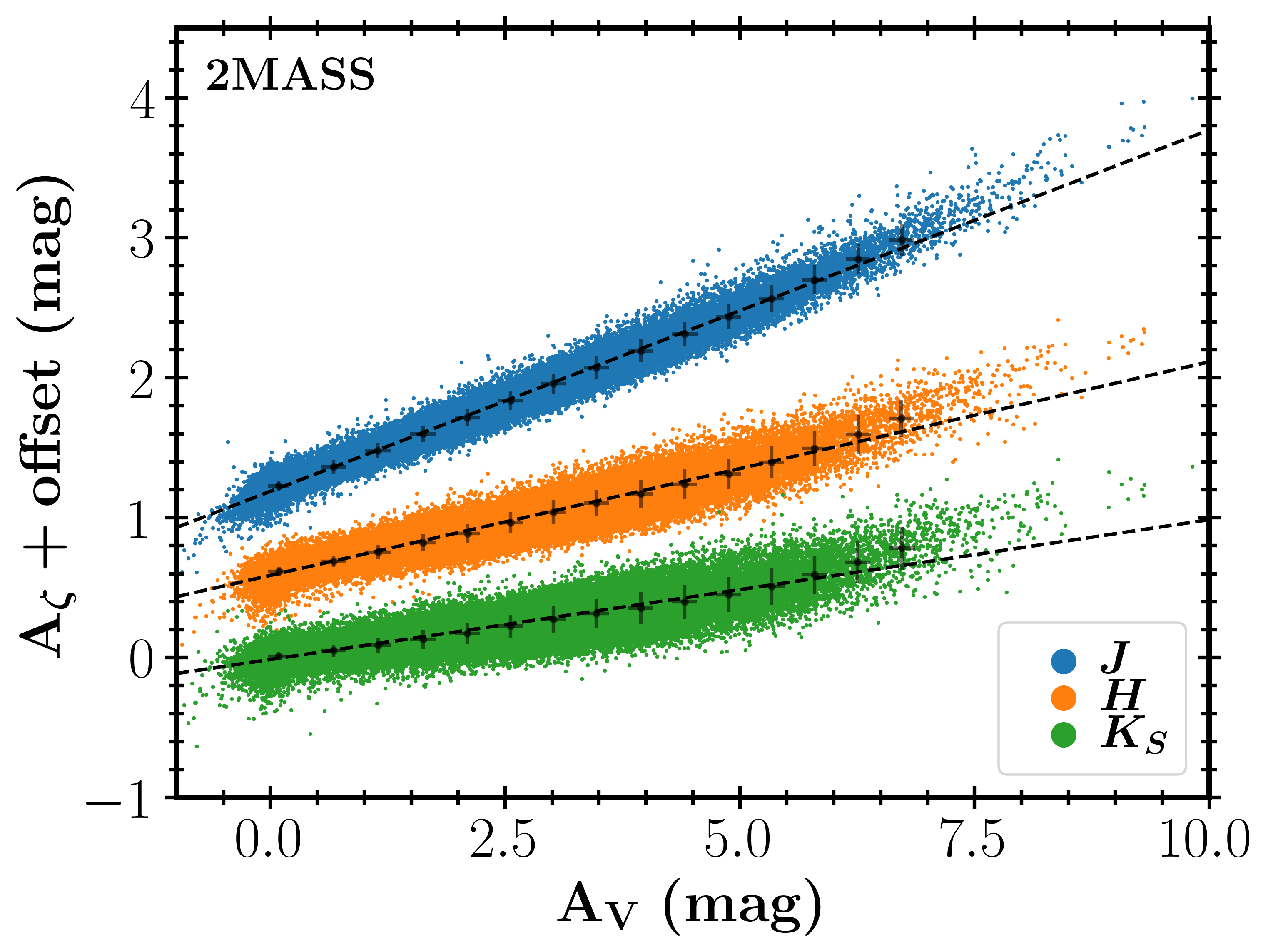}
    \includegraphics[width=0.88\columnwidth]{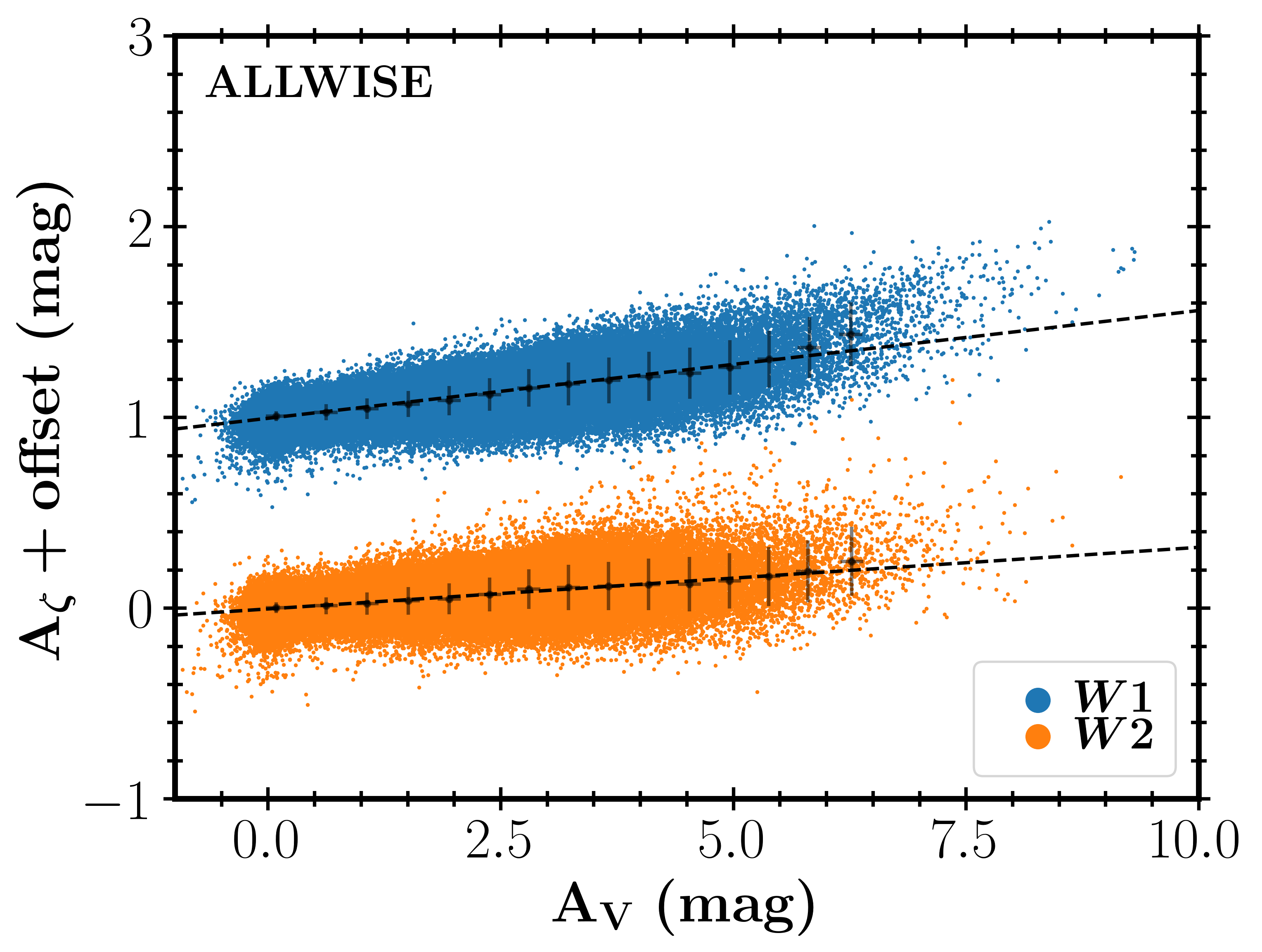}
\caption{
Similar to Fig.~\ref{fig:extinctionratioSyn}, but now for the following four photometric systems: (a) Gaia, (b) APASS, (c) 2MASS, and (d) ALLWISE, with extinction measurements based on observed photometry. For better visibility, extinction measurements for the \(H\), \(K_S\), and \(W1\) bands are shifted by 0.6, 1.2, and 1.0 mag, respectively. Larger extinction scatter is evident in the 2MASS and ALLWISE filters compared to other filters, due to their lower extinction sensitivity and larger photometric uncertainties. The extinction measurements (not shifted for any bands) and their uncertainties are available for download, while the derived extinction ratios and their uncertainties are provided in Table~\ref{tab:extinction}. See Section~\ref{sec: curve} for why $A_{G}$ does not extend to higher values as $A_{BP}$ and $A_{RP}$.}
\label{fig:extinctionratioObs}
\end{figure*}

Furthermore, we tested how  sensitive this anchor point was to the quality of the adopted parallax. Restricting the RC selection to a parallax precision ($\sigma_{\varpi}/\varpi$) better than 10\% (rather than 20\%) reduces the sample size from 8,426 to 7,719 stars and shifts $A_{BP}/A_{RP}$ only slightly, to $1.700 \pm 0.004$. This fractional variation of merely 0.3\% demonstrates that our adopted anchor point ($A_{BP}/A_{RP} = 1.694 \pm 0.004$) is  robust against parallax uncertainties.

Finally, we derived extinctions for all individual passbands using this anchor point, $A_{BP} / A_{RP} = 1.694 \pm 0.004$. Combining this ratio with the reddening $E(BP - RP) = A_{BP} - A_{RP}$, we calculated the individual extinctions $A_{BP}$ and $A_{RP}$ for all stars in the sample.  These base extinctions in $BP$ and $RP$ then allowed us to calculate the extinction $A_{\zeta}$ for any other passband $\zeta$ by utilizing the previously derived reddening values, $E(BP - \zeta)$ or $E(\zeta - RP)$, whichever was used in the $\texttt{XGBoost}$ training (see Section~\ref{samples}). Notably, this set of $A_{\zeta}$ includes $A_V$ for the $V$ band, which is used in the next Section. 
The uncertainties in $A_{BP}$ and $A_{RP}$ were derived by propagating the uncertainties in both the anchor ratio $A_{BP}/A_{RP}$ and the colour excess $E(BP - RP)$ using standard error-propagation techniques. Likewise, the uncertainties in $A_{\zeta}$ were estimated by combining the uncertainties in the corresponding reddening measurements, \mbox{$E(BP - \zeta)$} or $E(\zeta - RP)$, with the propagated uncertainty in the reference extinction, $A_{BP}$ or $A_{RP}$, depending on which band (i.e., $BP$ or $RP$) was adopted as the anchor during the training phase.

Since the anchor ratio $A_{BP}/A_{RP}$ was calibrated using a sample of RC-like stars, these extinction measurements are optimised for giant stars, which indeed comprise the majority of our APOGEE sample. We also performed a parallel analysis to derive the $A_{BP}/A_{RP}$ ratio for various main-sequence populations, binned in $T_{\rm eff}$ and [Fe/H] following the same scheme used to define our training and validation sets (see Section~\ref{samples}). However, the limited number of main-sequence stars—particularly those with high-precision \textit{Gaia} distances—results in substantially lower precision for the derived anchor ratios compared to the RC-like sample. Future data releases with larger samples of main-sequence stars, especially those with high-precision \textit{Gaia} parallaxes ($\sigma_\varpi/\varpi < 0.2$), will be valuable in refining these ratios across the entire Hertzsprung--Russell diagram.

We examined the dependence of the anchor point $A_{BP}/A_{RP}$ on $T_{\rm eff}$ using the Python package \href{https://github.com/vnohhf/extinction\_coefficient}{\texttt{extinction\_coefficient}} \citep{zhang2023}. We find that it varies typically within about $1-2\%$ relative to a typical RC-like star ($T_{\rm eff}=4750\,\mathrm{K}$) over the range $4000 \le T_{\rm eff} \le 6500\,\mathrm{K}$. Specifically, when this uncertainty is taken into account, we find that, for example, the median uncertainty in $A_{RP}$ increases only by a negligible amount of $0.003$–$0.007$~mag, which justifies the use of this anchor point to derive extinction from reddening for our sample.

\begin{table*}
\caption{Extinction ratios, $A_\zeta / A_V$, and extinction coefficients, $A_{\zeta}/E(B-V)$, and their uncertainties (see sections~\ref{sec: curve} and~\ref{sec: rv} for the methods). A machine-readable version of this table is available via the link provided in the Data Availability section.}
\label{tab:extinction}
\rowcolors{2}{gray!10}{gray!10}
\begin{tabular}{lcccccc}
\hline
\toprule
Filter & Effective~wavelength (\AA) & $A_{\zeta}/A_V$ & $\sigma_{A_{\zeta}/A{_V}}$ & $A_{\zeta}/E(B-V)$ & $\sigma_{A_{\zeta}/E(B-V)}$ & Synthetic~magnitude (Yes/No)\\
\midrule
APASS $B$ & 4299.2 & 1.179 & 0.014 & 3.804 & 0.045 & No \\
APASS $V$ & 5393.9 & 0.891 & 0.013 & 2.869 & 0.045 & No \\
APASS $g$ & 4639.3 & 1.118 & 0.008 & 3.574 & 0.014 & No \\
APASS $r$ & 6122.0 & 0.810 & 0.005 & 2.514 & 0.016 & No \\
APASS $i$ & 7438.9 & 0.618 & 0.001 & 1.945 & 0.006 & No \\
\midrule
GAIA $BP$ & 5035.8 & 0.958 & 0.003 & 3.110 & 0.008 & No \\
GAIA $RP$ & 7620.0 & 0.566 & 0.002 & 1.836 & 0.004 & No \\
GAIA $G$ & 5822.4 & 0.773 & 0.012 & 2.352 & 0.027 & No \\
\midrule
2MASS $J$ & 12350.0 & 0.258 & 0.002 & 0.810 & 0.007 & No \\
2MASS $H$ & 16620.0 & 0.153 & 0.003 & 0.478 & 0.009 & No \\
2MASS $K_\text{S}$ & 21590.0 & 0.100 & 0.004 & 0.304 & 0.009 & No \\
\midrule
WISE $W1$ & 33526.0 & 0.056 & 0.002 & 0.186 & 0.009 & No \\
WISE $W2$ & 46028.0 & 0.032 & 0.001 & 0.126 & 0.008 & No \\
\midrule
$U$ & 3551.1 & 1.604 & 0.012 & 5.172 & 0.017 & Yes \\
$B$ & 4361.9 & 1.303 & 0.004 & 4.149 & 0.004 & Yes \\
$V$ & 5467.6 & 1.000 & 0.001 & 3.146 & 0.003 & Yes \\
$R$ & 6695.8 & 0.767 & 0.002 & 2.464 & 0.004 & Yes \\
$I$ & 8568.9 & 0.585 & 0.002 & 1.811 & 0.003 & Yes \\
\midrule
HST $F435W$ & 4341.4 & 1.359 & 0.003 & 4.296 & 0.007 & Yes \\
HST $F606W$ & 5776.4 & 0.849 & 0.004 & 2.760 & 0.006 & Yes \\
HST $F814W$ & 8019.7 & 0.561 & 0.001 & 1.767 & 0.003 & Yes \\
\midrule
PS1 $g$ & 4810.2 & 1.143 & 0.003 & 3.642 & 0.004 & Yes \\
PS1 $r$ & 6155.5 & 0.854 & 0.001 & 2.680 & 0.003 & Yes \\
PS1 $i$ & 7503.0 & 0.644 & 0.001 & 1.998 & 0.002 & Yes \\
PS1 $y$ & 9613.6 & 0.423 & 0.004 & 1.289 & 0.008 & Yes \\
PS1 $z$ & 8668.4 & 0.514 & 0.003 & 1.568 & 0.005 & Yes \\
\midrule
SDSS $u$ & 3608.0 & 1.627 & 0.013 & 5.211 & 0.029 & Yes \\
SDSS $g$ & 4671.8 & 1.205 & 0.002 & 3.812 & 0.004 & Yes \\
SDSS $r$ & 6141.1 & 0.870 & 0.001 & 2.712 & 0.003 & Yes \\
SDSS $i$ & 7457.9 & 0.646 & 0.001 & 2.005 & 0.003 & Yes \\
SDSS $z$ & 8922.8 & 0.481 & 0.003 & 1.476 & 0.005 & Yes \\
\midrule
SkyMapper $v$ & 3878.7 & 1.468 & 0.006 & 4.660 & 0.009 & Yes \\
SkyMapper $g$ & 5016.1 & 1.073 & 0.002 & 3.425 & 0.006 & Yes \\
SkyMapper $r$ & 6076.8 & 0.852 & 0.001 & 2.688 & 0.003 & Yes \\
SkyMapper $i$ & 7732.8 & 0.611 & 0.001 & 1.896 & 0.003 & Yes \\
SkyMapper $z$ & 9120.3 & 0.454 & 0.003 & 1.400 & 0.006 & Yes \\
\midrule
Strömgren $v$ & 4105.3 & 1.430 & 0.004 & 4.433 & 0.013 & Yes \\
Strömgren $b$ & 4693.8 & 1.291 & 0.004 & 4.076 & 0.006 & Yes \\
Strömgren $y$ & 5500.9 & 1.050 & 0.004 & 3.249 & 0.007 & Yes \\
\bottomrule
\end{tabular}
\end{table*}

\section{Precision and Accuracy of Extinction Measurements}\label{sec: precision_accuracy}
In this section, we validate the precision and accuracy of our derived extinction measurements. We compare our results with three established, independent studies: the 3D dust maps of \texttt{Bayestar19} \citep{green2019a}, the \texttt{StarHorse} catalog derived from isochrone fitting \citep{queiroz2023}, and the SED-based extinctions obtained through the \texttt{SEDEX} pipeline \citep{yu2021,yu2023}. It is important to note the differences in methodology: the $A_V$ values from \texttt{Bayestar19} are photometry-based, calculated from their $E(B-V)$ values assuming $R_V = 3.1$. In contrast, the $A_V$ values derived in this work, as well as those from \texttt{StarHorse} and \texttt{SEDEX}, rely on spectroscopic stellar parameters (from APOGEE DR17 for \texttt{StarHorse} and DR19 for \texttt{SEDEX}).

 \begin{figure*}
	\includegraphics[width=0.63\textwidth, trim=0 0 0 0, clip]{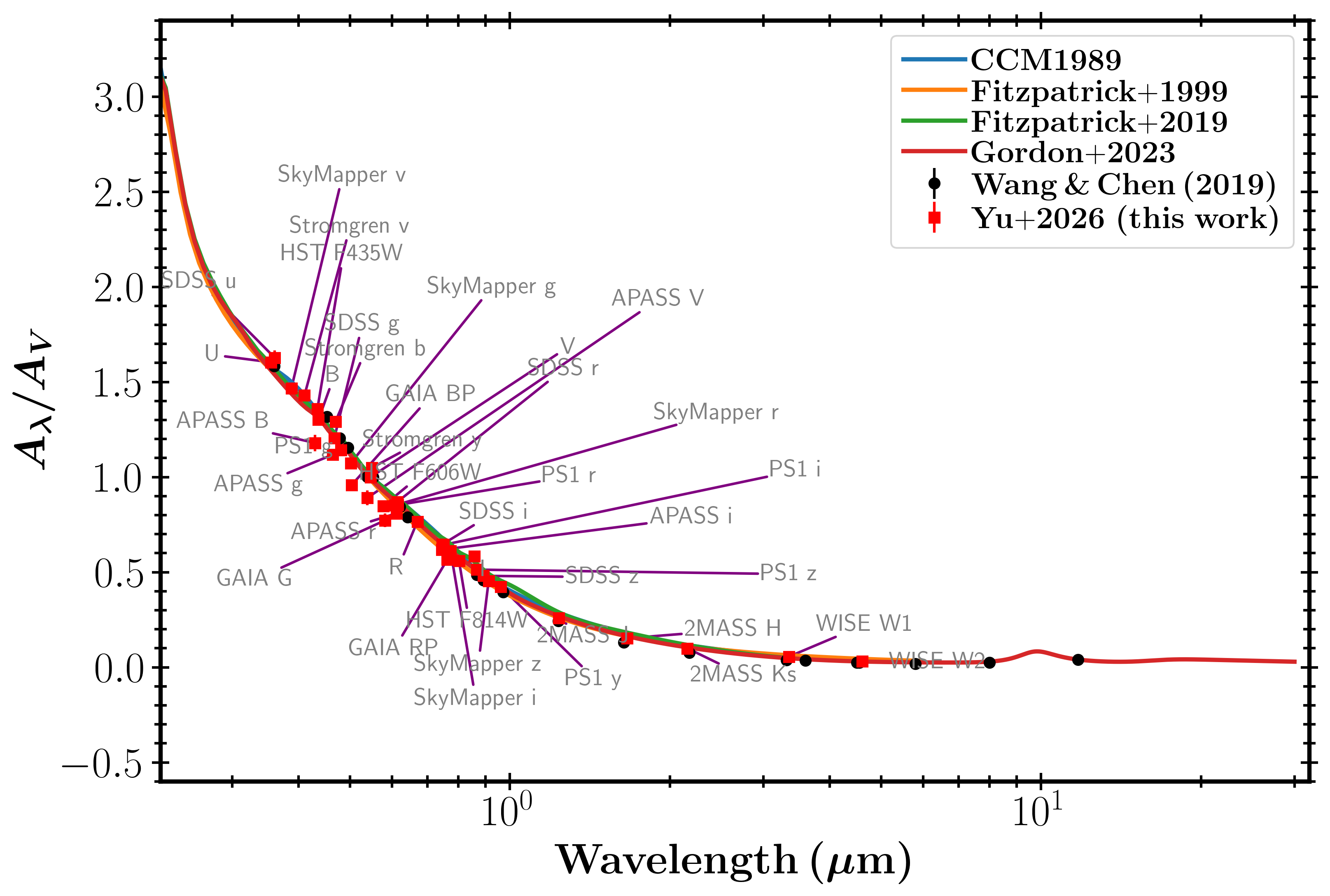}\\
    \includegraphics[width=0.63\textwidth, trim=0 0 0 0, clip]{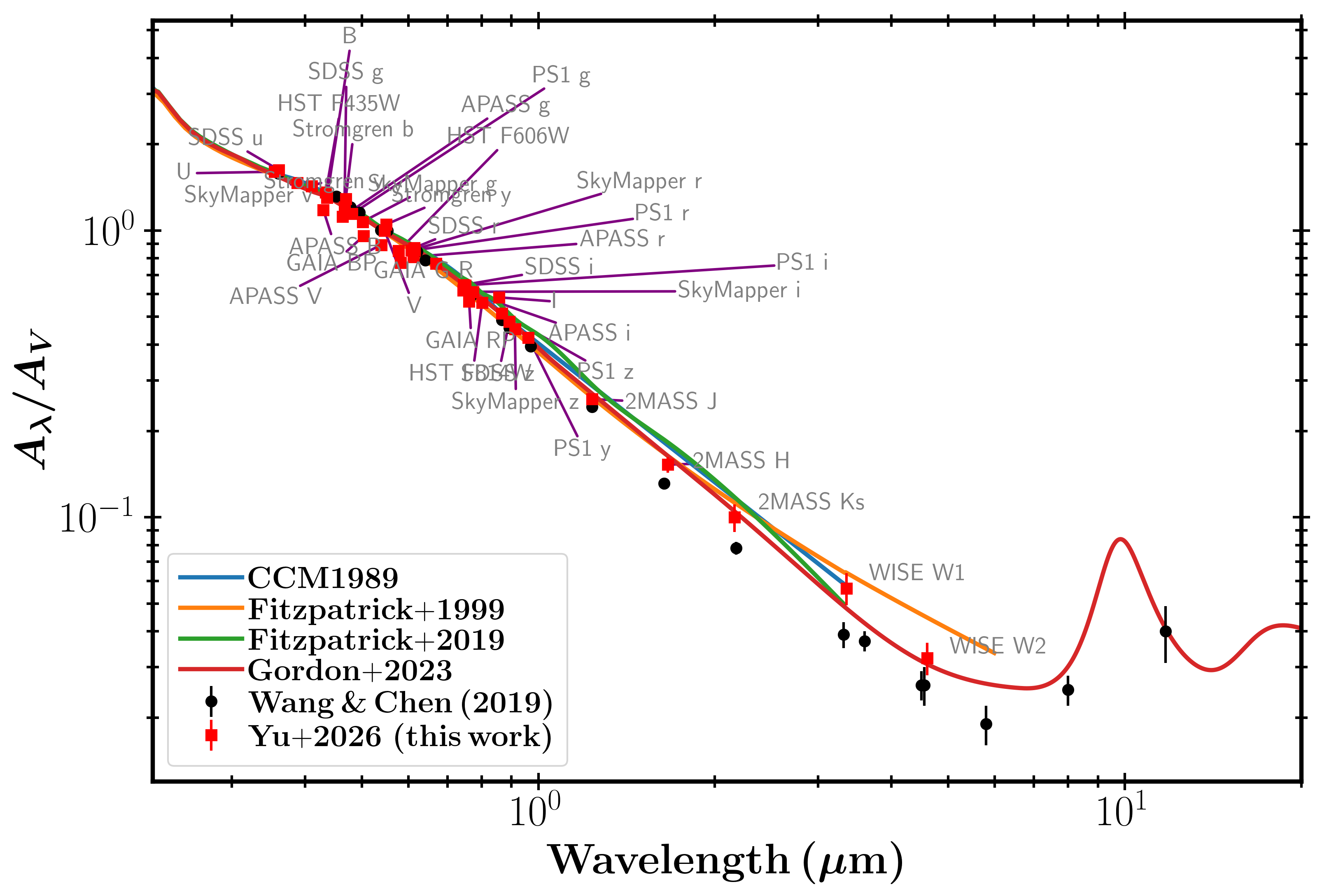}\\
\caption{
A new extinction curve derived in this work, where extinctions (\(A_\zeta\) and \(A_V\)) are calculated for entire passbands rather than static wavelengths (e.g., effective wavelengths of individual passbands). \textbf{Upper panel:} Comparison of the extinction ratios \(A_{\zeta}/A_V\) derived in this work (red squares) with those from literature extinction curves assuming \(R_V=3.1\): \citet[][blue line]{cardelli1989}, \citet[][orange line]{fitzpatrick1999}, \citet[][green line]{fitzpatrick2019}, and \citet[][red line]{gordon2023}. Black circles represent extinction ratios from \citet{wang2019}. The extinction ratios derived in this work are shown with their 3\(\sigma\) uncertainties, and the extinction values for each filter are labeled. \textbf{Lower panel:}  Similar to the top panel, but now showing the extinction ratios on a logarithmic scale to better illustrate differences at longer wavelengths.}
\label{fig:extinctionlaw}
\end{figure*}

The upper panel of Fig.~\ref{fig:avprecision} demonstrates that, while all methods show general agreement up to $A_V \simeq 8$, the scatter between them varies significantly (see Fig.s~\ref{fig:appendix_fig1}, \ref{fig:appendix_fig2}, and \ref{fig:appendix_fig3} for detailed comparisons between our $V$-band extinction measurements and those from \texttt{Bayestar19}, \texttt{StarHorse}, and \texttt{SEDEX}, respectively). Extinctions derived in this work align closely with those from \texttt{StarHorse} and show the tightest correlation with \texttt{SEDEX}. Conversely, the \texttt{Bayestar19} values exhibit the largest scatter, particularly at high extinction values. The differences in the scatter are attributed to the differences in the data used: methods relying on spectroscopic data (this work, \texttt{StarHorse}, \texttt{SEDEX}) typically provide higher extinction precision than photometry-based methods (\texttt{Bayestar19}). Regarding the overall accuracy of our measurements, we fitted a linear model comparing our $A_V$ values with those from the literature. This analysis revealed that our extinction values are globally $17.3\%$ larger than \texttt{Bayestar19}, $11.3\%$ larger than \texttt{SEDEX}, and $3.0\%$ smaller than \texttt{StarHorse}. 

We note that the zero point of our extinction measurements, derived using the star-pair method, should be superior to those obtained using \texttt{StarHorse} and \texttt{SEDEX}, by design.
If biases exist in the APOGEE stellar parameters, they do not necessarily propagate into our intrinsic color estimates. Because $\texttt{XGBoost}$ is a non-parametric model capable of capturing complex, non-linear dependencies, it can effectively model the systematic biases of stellar parameters as a function of the parameters themselves. As long as these biases remain consistent for stars with similar properties (i.e., \teff, \logg, \feh, and \afe) regardless of their sky position, the model implicitly accounts for them during the mapping from observed parameters to intrinsic colors.
On the other hand, any biases in the APOGEE stellar parameters can translate into extinction measurements derived from $\texttt{StarHorse}$ and $\texttt{SEDEX}$ because those methods must match observed and model quantities, both of which could introduce biases in the final extinction values. Therefore, based on this robust methodology, we did not correct our extinction scale for subsequent analysis.

The internal precision of our $A_V$ measurements is assessed from the dispersion of $A_V$ values for 802 nearby stars ($d < 70\ \text{pc}$); this sample lies within the Local Bubble, a region known to have virtually zero reddening \citep{leroy1993, lallement2014}. Because these stars are expected to experience negligible extinction, the dispersion is thus dominated by random noise in the $A_V$ measurements. The lower panel of Fig.~\ref{fig:avprecision} compares the $A_V$ distributions for this nearby sample derived from this work, \texttt{StarHorse}, and \texttt{SEDEX}. Extinctions from this work (blue) exhibit the least scatter among the three methods, yielding a $1\sigma$ uncertainty of $0.042\ \text{mag}$. This precision is $2.3$ times better than that of \texttt{SEDEX} (purple) and $8.7$ times better than that of \texttt{StarHorse} (green). Furthermore, the median $A_V$ from this work is $0.001\ \text{mag}$, which is consistent with the expected negligible extinction for these stars \citep{leroy1993, lallement2014}. Crucially, the median uncertainty of our $A_V$ measurements ($0.03\ \text{mag}$) is consistent with the measured $A_V$ dispersion ($0.042\ \text{mag}$), suggesting that our internal uncertainty estimates are reliable.

We did not use \textit{Gaia} distances to derive extinction values in this work. The main reason is that our method allows extinction measurements to be obtained for stars whose \textit{Gaia} distances are either unavailable or have large uncertainties. As shown in Fig.~\ref{fig:avprecision}, a number of stars have extinction values exceeding $A_V = 6$, a regime where \textit{Gaia} parallaxes become much less precise or are unavailable. In addition, if $A_V$ were derived from distance using the relation $A_V = V - M_V - 5\log d + 5$, then reproducing the typical extinction uncertainty of $0.03~\mathrm{mag}$ reported in this work would require a distance precision of about $6.5\%$. Such a level of distance precision is generally difficult to achieve for highly reddened stars in the Galactic plane.  Therefore, to derive extinction values across both low- and high-extinction regions of the Milky Way, we adopted a homogeneous approach that does not rely on \textit{Gaia} distances.

\begin{figure}
	\includegraphics[width=\columnwidth]{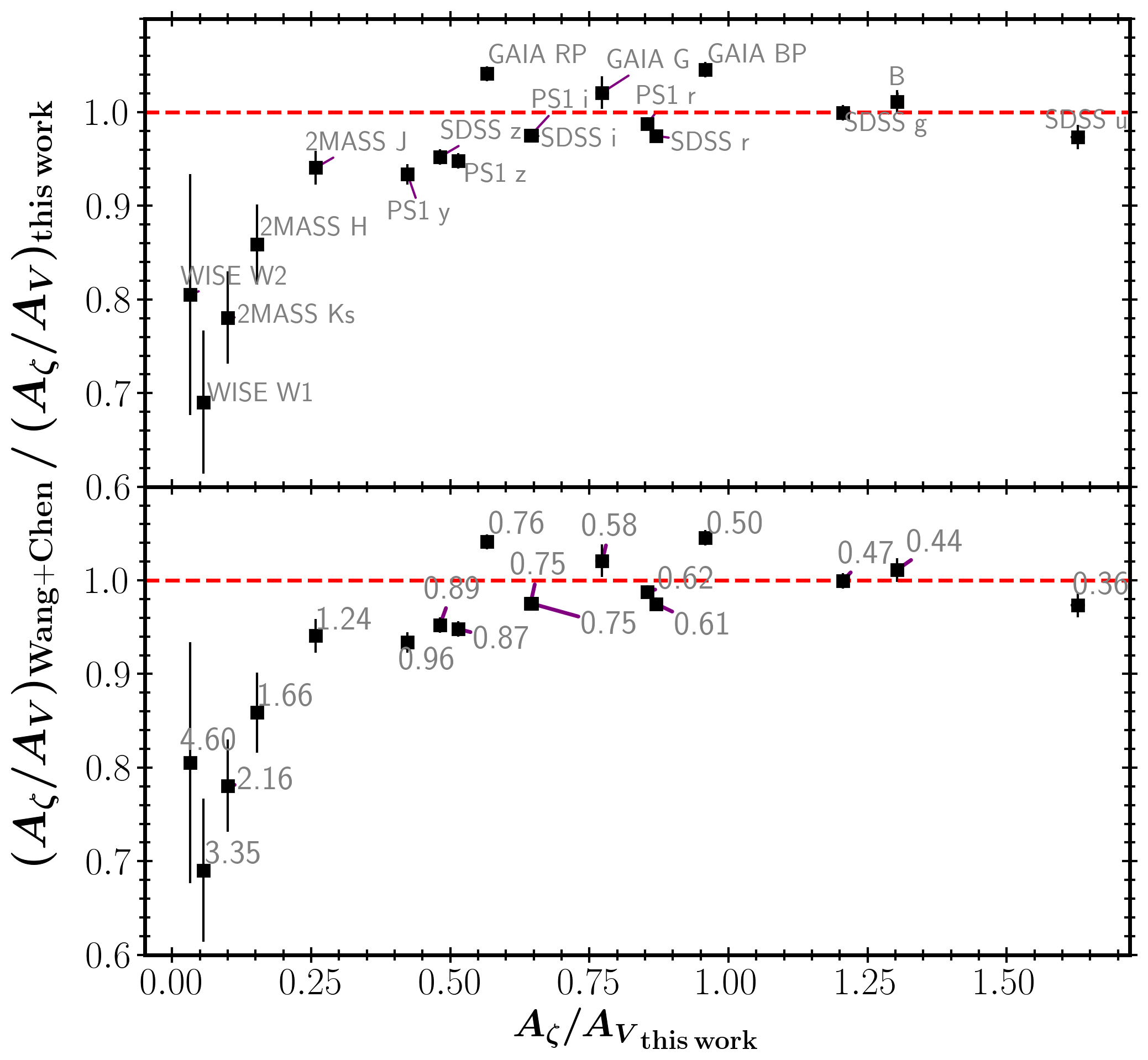}
\caption{Comparison of extinction ratios derived in this work with those from \citet{wang2019}.
\textbf{Upper panel:} Each point corresponds to an individual filter, with labels indicating the photometric band. The horizontal dashed line marks a perfect agreement between the two measurements.
\textbf{Lower panel:} Same ratio as shown above, with numbers indicating the effective wavelengths (in $\mu$m) of the corresponding filters, adopted from the \href{http://svo2.cab.inta-csic.es/svo/theory/fps3/index.php?mode=browse}{SVO} Filter Profile Service. Deviations from unity increase toward longer wavelengths, particularly in the near- and mid-infrared. \label{fig:wangextinctionlaw}
}
\end{figure}

\section{A New Extinction Curve}\label{sec: curve} 
With our passband extinction measurements for 39 filters, we proceeded to derive extinction ratios for building a new extinction curve. Fig.~\ref{fig:extinctionratioSyn} illustrates the relationship between $A_V$ and the derived extinction $A_{\zeta}$ for 26 filters across six photometric systems: (a) SDSS, (b) Pan-STARRS, (c) SkyMapper, (d) Johnson-Kron-Cousins, (e) HST, and (f) Strömgren. These extinction measurements were derived from synthetic magnitudes based on $\textit{Gaia}$ BP/RP spectra; with the exception of SkyMapper, all magnitudes were standardized by the \textit{Gaia} collaboration \citep{gaiacollaboration2023}. 

In Fig.~\ref{fig:extinctionratioSyn}, each panel shows the extinction values for individual filters for a given photometric system, revealing strong linear correlations between $A_V$ and $A_{\zeta}$. That being said, the detectable curvatures previously observed in reddening-to-reddening relationships (which tend toward quadratic at high reddening; see Fig.~\ref{fig:reddeningcomp}) do not clearly propagate into the extinction-to-extinction relationships, which remain essentially linear. To calculate the extinction ratio $A_{\zeta}/A_V$ for each filter $\zeta$, we fitted a linear model to the binned averages of $A_{\zeta}$ and $A_V$, represented by crosses in the figures, while accounting for the standard error in each bin. We find that these ratios show no significant correlation with either \teff\ or $A_V$, which is consistent with the minimal dispersion observed around the linear trends in Fig.~\ref{fig:extinctionratioSyn}. The resulting best-fit extinction ratios and their uncertainties are listed in Table~\ref{tab:extinction}.  

Extinction measurements for blue filters, such as the SDSS $u$ band, were available for fewer stars due to our quality cuts on synthetic magnitudes, which required the associated flux-to-flux error ratio to exceed 30. Because the stars analyzed in this work are predominantly cool red giants, with stars of $T_{\text{eff}} > 8000\ \text{K}$ excluded, bluer bands like the SDSS $u$ band were more affected by lower fluxes in these filters, leading to fewer stars meeting the quality criteria. 

Similar to the analysis of the synthetic magnitudes for the 6 systems, Fig.~\ref{fig:extinctionratioObs} presents extinction measurements based on observed magnitudes for four photometric systems: (a) Gaia, (b) APASS, (c) 2MASS, and (d) ALLWISE. The derived extinction ratios and their associated uncertainties were calculated using the same method and are also provided in Table~\ref{tab:extinction}.

\begin{figure}
	\includegraphics[width=\columnwidth]{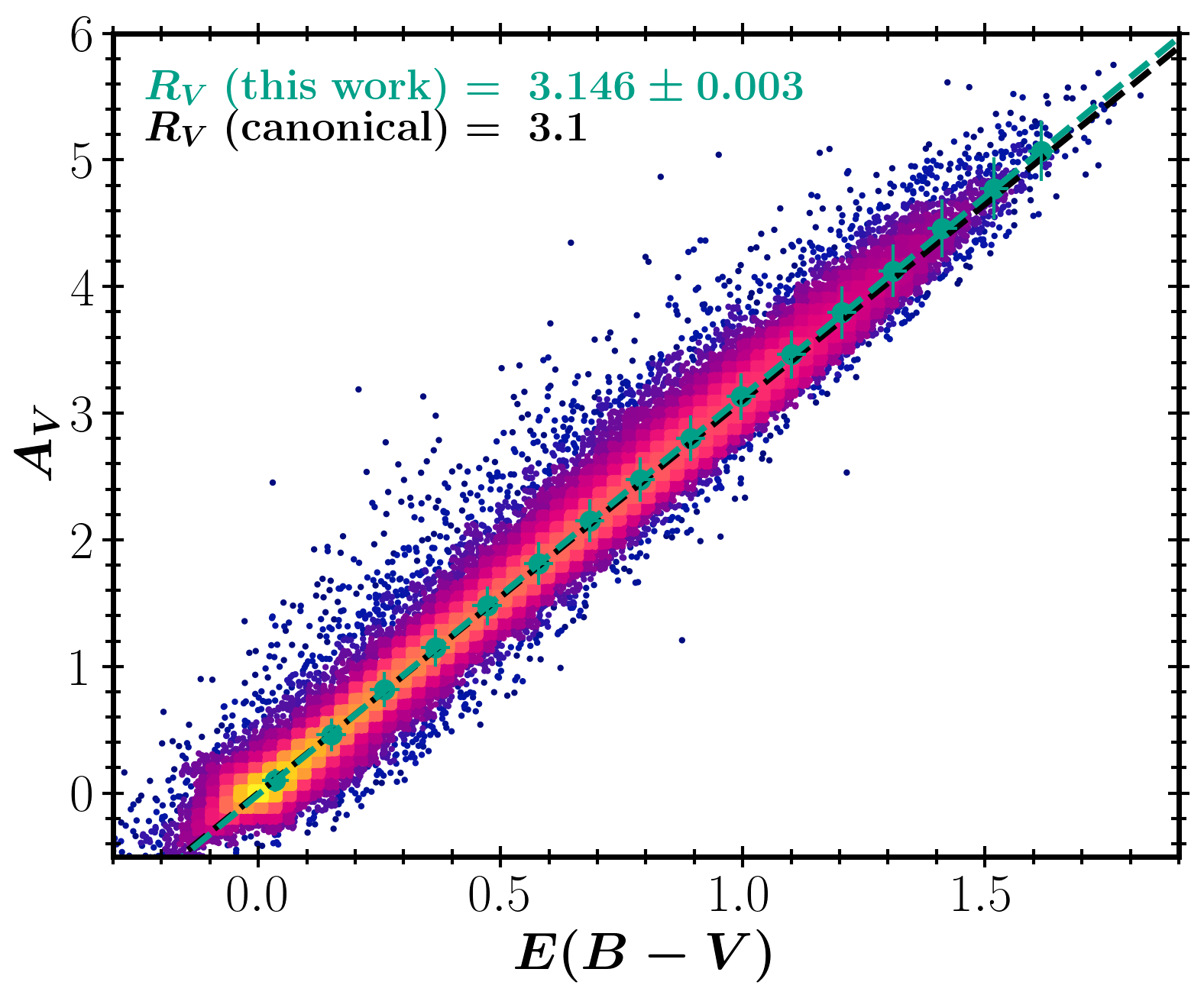}
\caption{Relationship between \(A_V\) and \(E(B-V)\), color-coded by star number density, with yellow indicating higher density and blue indicating lower density. The green points represent binned averages with their corresponding standard errors. The best-fit relationship, shown by the green dashed line, has a slope of \(R_V = 3.146 \pm 0.003\), while the canonical relationship with a slope of \(R_V = 3.1\) is indicated by the black dashed line.}
\label{fig:rv}
\end{figure}

\begin{figure*}
	\centering
	\includegraphics[width=0.82\textwidth]{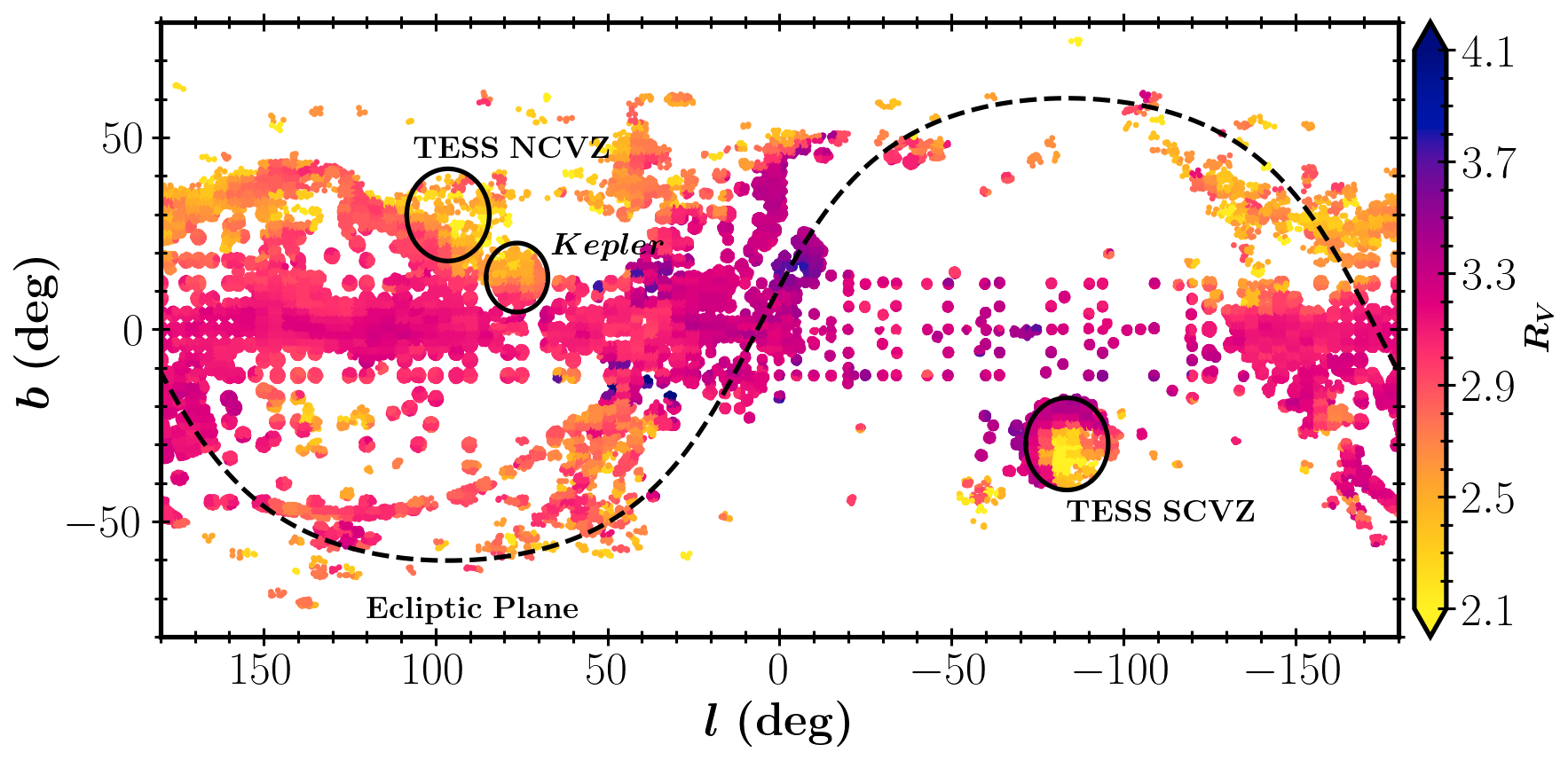}
	\caption{
	Sky map of the extinction coefficient \(R_V = A_V / E(B-V)\) for the APOGEE DR19 footprint in Galactic coordinates (\(l, b\)). The color bar indicates the average \(R_V\) within each bin, with a resolution of 90 bins in both the \(l\) and \(b\) dimensions. Bins containing fewer than four stars and stars with \(E(B-V) < 0.05\) or \(A_V < 0.1\) are excluded to ensure reliable measurements. The color scale is truncated to the range \(2.1 \leq R_V \leq 4.1\), corresponding to the interval in which the statistics are well sampled; values outside this range are clipped to the color-bar limits. Within this range, higher \(R_V\) values are predominantly found near the Galactic plane, while lower values are more common at higher Galactic latitudes. The black dashed line denotes the ecliptic plane, along which the $K2$ fields are situated. The footprints of the \textit{Kepler} field and the TESS northern and southern Continuous Viewing Zones (CVZs) are also indicated, all of which demonstrate significant spatial variations in $R_V$.}
	\label{fig:rvSky}
\end{figure*}

We note that extinction measurements for the $\textit{Gaia}$ $G$-band were systematically unavailable for stars with higher extinction values. 
This is a consequence of our SED fitting process designed to filter outlier photometry: at high extinction, the monochromatic extinction predicted by a standard extinction curve deviates significantly from the passband-integrated extinction reflected in the observed $G$-band magnitudes \citep[e.g.][]{casagrande2014,wang2019,zhang2023}. Because these measurements appeared as outliers relative to the model, they were removed during the quality-control phase through the SED fitting. 
Despite this limitation, $A_G$ still exhibits a clear linear relationship with $A_V$ when $A_V \lesssim 4\ \text{mag}$, suggesting that the differences between monochromatic and passband extinction are negligible in this regime. 

Although the APASS photometric system is similar to the Johnson-Kron-Cousins system for the $B$ and $V$ filters and to the SDSS system for the $g$, $r$, and $i$ filters, we still derived extinction ratios specifically for the APASS filters. This approach ensured that the derived extinction ratios aligned with the photometric zero points unique to the APASS system. Extinction measurements for the 2MASS and ALLWISE filters show larger uncertainties compared to other filters. This is primarily due to their lower sensitivity to extinction and higher intrinsic photometric uncertainties, which contributes to the greater scatter observed in their extinction measurements. 

Fig.~\ref{fig:extinctionlaw} presents the extinction curve derived in this work (39 filters) and compares it with five literature extinction curves assuming $R_V=3.1$: \citet{cardelli1989}, \citet{fitzpatrick1999}, \citet{fitzpatrick2019}, \citet{wang2019} and \citet{gordon2023}. Our derived curve exhibits overall consistency with the literature, particularly at shorter wavelengths in the ultraviolet (UV) and optical bands. However, at longer wavelengths, particularly in the infrared (IR) region (e.g., WISE and 2MASS bands), discrepancies become more pronounced, as shown in the lower panel of Fig.~\ref{fig:extinctionlaw}. Notably, our extinction curve aligns closely with \citet{gordon2023}, but it exhibits systematic differences from \citet{wang2019}.

Fig.~\ref{fig:wangextinctionlaw} compares our extinction curve with that of \citet{wang2019}. Overall consistency is observed across optical wavelengths, with most ratios clustering around unity; however, systematic deviations emerge toward longer wavelengths, where \citet{wang2019} yield systematically lower extinction ratios. The lower panel shows that this departure from unity increases from wavelengths of $\sim700\,\mathrm{nm}$ into the near- and mid-infrared, particularly for the 2MASS and WISE bands. These discrepancies primarily reflect methodological differences: our extinction ratios are derived from passband-integrated measurements using observed and synthetic photometry, whereas \citet{wang2019} report extinction ratios evaluated at fixed effective wavelengths. In addition, extinction estimates in the near- and mid-infrared are intrinsically less precise owing to the reduced sensitivity and broader bandpasses of these filters, which further contribute to the observed scatter.

In conclusion, the broadband extinction ratios derived in this work are optimal for correcting extinctions for individual filters, ensuring consistency with passband-integrated observations. Finally, for applications that require the ratios of $A_{\zeta}/A_V$ and $A_{\zeta}/E(B-V)$ tailored to specific spectral types \citep[e.g.,][]{zhang2023} or evolutionary stages \citep[e.g.,][]{wang2025b}, users can use the extinction measurements provided in our catalog to derive custom ratios beyond the average values reported in Table~\ref{tab:extinction}.

\section{Extinction Coefficient $R_V$ Across the APOGEE DR19 Sky Footprint}\label{sec: rv}
The consistency between our newly derived extinction curve and the literature results suggests that the average $R_V$ for our stellar sample is $\sim$$3.1$. To quantify this, we analyzed the relationship between $A_V$ and $E(B-V)$ as shown in Fig.~\ref{fig:rv}. A linear fit to the binned averages (represented by green filled circles) yields a slope of $3.146 \pm 0.003$, aligning closely with the average value of 3.1 in the Milky Way (indicated by the black dashed line, \citealt{cardelli1989, schlafly2011,zhanggreen2025}). 
Notably, the scatter around this best-fit relation shown Fig.~\ref{fig:rv} is $0.09\ \text{mag}$, which exceeds our typical $A_V$ uncertainty of $0.03\ \text{mag}$. This discrepancy implies that the observed dispersion is driven by intrinsic variations in the extinction curve rather than by random measurement errors. 

It is well established that such variations in $R_V$ can be typically attributed to differences in dust grain size distributions, chemical compositions, and local radiation environments \citep{schlafly2016, green2019a, zhangruoyi2023}. Fig.~\ref{fig:rvSky} illustrates the spatial distribution of the extinction coefficient $R_V = A_V / E(B-V)$ across the APOGEE DR19 footprint. We observe that higher $R_V$ values ($\sim$4.1) are predominantly concentrated near the Galactic plane, whereas lower values ($\sim$2.1) characterize the higher Galactic latitudes. It is noteworthy that the large variation of $R_V$ suggests that the initial photometric filtering—which relied on SED fitting as detailed in Section~\ref{sec: sed}—does not significantly bias the sample by removing stars located behind "non-standard" dust. The spatial trends we see are in agreement with previous studies \citep[e.g.,][]{schlafly2016, zhangruoyi2023, zhanggreen2025, green2025}.

The observed spatial gradient in $R_V$ reflects the heterogeneous physical conditions of the interstellar medium (ISM, \citealt{draine2003}). The higher $R_V$ values found within the Galactic plane point to environments dominated by larger dust grains, a characteristic feature of the dense molecular clouds associated with active star formation. In these high-density regions, processes such as grain coagulation and mantle accretion facilitate the growth of larger particles, thereby increasing $R_V$ \citep{ossenkopf1993}. In contrast, the lower $R_V$ values observed at higher latitudes are indicative of the diffuse ISM. In these less-shielded, lower-density environments, small grains are more prevalent due to fragmentation from interstellar shocks or erosion by the interstellar radiation field \citep{jones1996}.

The spatial distribution of $R_V$ across key space-mission footprints—namely the \textit{Kepler} field and the TESS northern and southern Continuous Viewing Zones (CVZs)—exhibits pronounced small-scale variations that have important implications for precision stellar astrophysics. These regions are prioritized for long-duration photometric monitoring to enable exoplanet characterization and asteroseismic analysis. Our results reveal that $R_V$ is markedly non-uniform across all three fields (with stars in the Large Magellanic Cloud removed), reflecting the intrinsically inhomogeneous structure of the intervening interstellar medium. Such spatial variations in $R_V$ can introduce significant systematic biases in fundamental stellar parameters, including luminosities and radii, when \textit{Gaia} parallaxes are combined with broadband photometry under the assumption of a standard Galactic average ($R_V = 3.1$). Therefore, accurate, position-dependent $R_V$ corrections are essential for achieving the precision required in modern stellar studies.

Fig.~\ref{fig:appendix_fig4} shows the comparison of our measured $R_V$ estimates with those predicted by the 3D $R_V$ map of \citet{zhanggreen2025}, which is based on \textit{Gaia} BP/RP spectra for stars extending beyond 5 kpc. To perform this comparison, we obtained predicted $R_V$ values by interpolating their integrated 3D $R_V$ map at the positions of individual stars using their sky coordinates and \textit{Gaia} distances. As shown in Appendix Fig.~\ref{fig:appendix_fig4}, overall, we find reasonably good agreement between the two estimates, with a scatter of $\sim 0.8$ (see the similar validation shown in Fig.~S5 of the supplementary material in \citet{zhanggreen2025}). Some systematic differences are expected because the $R_V$ values in Zhang \& Green (2025) are not defined using the classical Johnson $B$ and $V$ bandpasses (whereas ours are), which may lead to a different $R_V$ scale compared to our definition.

We estimate the systematic uncertainty in our $R_V$ values to be $\sim$3--6\%. Our extinction curve is anchored using the $A_{BP}/A_{RP}$ ratio of RC-like stars, rather than Gaia parallaxes. This choice is motivated by its suitability for high-extinction stars ($A_V \gtrsim 6~\mathrm{mag}$), for which parallax-based distances are uncertain. Accounting for the dependence of this ratio on effective temperature, we show in Section~\ref{sec: ext} that $A_{BP}/A_{RP}$ varies at the 1--2\% level across our sample. Using the CCM89 extinction curve, this corresponds to a 3--6\% variation in $R_V$, indicating that the resulting extinction curve is robust against such systematic effects.

In addition to $R_V$, we derived extinction coefficients (i.e., total-to-selective extinction ratios), $A_\zeta / E(B-V)$, for additional 38 filters using the same linear regression method employed for $R_V$. These measured coefficients, along with their respective uncertainties, are detailed in Table~\ref{tab:extinction}. This suite of coefficients provides a toolset for applying extinction corrections across a broad range of photometric systems, particularly when utilizing reddening values (e.g., $E(B-V)$) derived from 3D dust maps. In addition, our publicly available reddening measurements enable users to derive reddening ratios for arbitrary colour combinations among the 39 filters considered (e.g., $E(BP-RP)/E(B-V)$).

\section{Conclusion}
High-precision reddening and extinction measurements are of fundamental importance to a wide range of stellar and interstellar studies. In particular, they are essential for applying accurate extinction corrections when deriving precise stellar parameters from \textit{Gaia} parallaxes. The primary objective of this study was to establish a high-fidelity catalog of reddening and extinction measurements for the APOGEE DR19 stars. Using these data, we derived extinction ratios ($A_{\zeta}/A_V$) and extinction coefficients ($A_{\zeta}/E(B-V)$) specifically optimized for broadband photometry.

To achieve this, we leveraged high-quality spectroscopic parameters from APOGEE DR19 and standardized synthetic magnitudes from \textit{Gaia} BP/RP spectra to estimate intrinsic colors. This was accomplished using an \texttt{XGBoost} regression model trained on a sample of stars with negligible reddening ($E(B-V)_{\rm SFD} < 0.02$). By comparing these predicted intrinsic colors with observed measurements, we derived precise reddening values. These color excesses were subsequently converted into absolute extinctions, anchored by a precisely determined ratio of $A_{BP} / A_{RP} = 1.694 \pm 0.004$ obtained from a sample of RC-like stars (Fig.~\ref{fig:anchor}).

The resulting catalog provides high-fidelity extinction and reddening estimates for 39 filters across 10 distinct photometric systems (publicly available for download at \url{https://doi.org/10.5281/zenodo.18239586}) \citep{yu_2026_18239586}. This catalog enables the derivation of precise extinction ratios ($A_{\zeta}/A_V$), extinction-to-reddening ratios ($A_{\zeta}/E(B-V)$), and reddening ratios (e.g., $E(BP-RP)/E(B-V)$) for any combination of the following passbands:
\begin{itemize}
    \item \textbf{Gaia:} $G$, $BP$, $RP$
    \item \textbf{Pan-STARRS1:} $g, r, i, z, y$
    \item \textbf{SDSS:} $u, g, r, i, z$
    \item \textbf{APASS:} $B, V, g, r, i$
    \item \textbf{Johnson-Kron-Cousins:} $U, B, V, R, I$
    \item \textbf{SkyMapper:} $v, g, r, i, z$
    \item \textbf{Strömgren:} $v, b, y$
    \item \textbf{HST:} $F435W, F606W, F814W$
    \item \textbf{2MASS:} $J, H, K_S$
    \item \textbf{ALLWISE:} $W1, W2$
\end{itemize}

Compared with extinction measurements from existing catalogs and methods—namely \texttt{Bayestar19} \citep{green2019a}, \texttt{StarHorse} \citep{queiroz2023}, and \texttt{SEDEX} \citep{yu2021, yu2023}---our results demonstrate superior precision, achieving a typical $A_V$ uncertainty of $\sim$0.03~mag. The newly derived extinction curve shows good agreement with established literature \citep{cardelli1989, fitzpatrick1999, fitzpatrick2019, gordon2023, wang2019} at shorter wavelengths in the UV and optical regimes (Fig.~\ref{fig:extinctionlaw}). However, notable discrepancies emerge at longer wavelengths ($\lambda > 700$~nm), particularly in the near- and mid-infrared (e.g., 2MASS and \textit{WISE} bands). These differences likely arise from the varying treatments of extinction ratios across studies—specifically the distinction between broadband-integrated and monochromatic ratios. While the newly derived extinction curve characterizes an average $R_V = 3.146 \pm 0.003$ (Fig.~\ref{fig:rv}), we resolve significant spatial variations in $R_V$, with higher values concentrated near the Galactic plane and lower values at higher Galactic latitudes (Fig.~\ref{fig:rvSky}).

The methodology developed in this work has yielded a precise reddening and extinction catalog that will serve as a foundation for several future applications. As part of a forthcoming series of studies, the superior precision of these measurements will enable a revision of the $\nu_{\rm{max}}$ asteroseismic scaling relation \citep{hekker2020} and provide critical anchor points for the absolute calibration of differential reddening in open clusters \citep{kalup2026}, leading to more reliable age determinations via isochrone fitting. Furthermore, by exploiting these high-precision measurements across an extensive number of lines of sight, our results offer a robust mechanism to reconcile existing heterogeneous dust maps \citep[e.g.,][]{drimmel2003, marshall2006, sale2014, green2019a, lallement2022, wang2025c} into a single, all-sky homogeneous extinction model. This unified framework will provide a consistent view of the dust distribution within the Milky Way, facilitating more accurate studies of Galactic structure and stellar populations.

\section*{Acknowledgements}
We thank the referee for their helpful comments and suggestions, which have helped improve the manuscript. RD is supported in part by the Italian Space Agency (ASI) through contract ASI-INAF 2025-10-HH.0 to the National Institute for Astrophysics (INAF). D.S. is supported by the Australian Research Council (DP250104267). SK acknowledges support from the European Union's Horizon 2020 research and innovation program under the GaiaUnlimited project (grant agreement No 101004110).

Funding for the Sloan Digital Sky Survey V has been provided by the Alfred P. Sloan Foundation, the Heising-Simons Foundation, the National Science Foundation, and the Participating Institutions. SDSS acknowledges support and resources from the Center for High-Performance Computing at the University of Utah. SDSS telescopes are located at Apache Point Observatory, funded by the Astrophysical Research Consortium and operated by New Mexico State University, and at Las Campanas Observatory, operated by the Carnegie Institution for Science. The SDSS web site is \url{www.sdss.org}.

SDSS is managed by the Astrophysical Research Consortium for the Participating Institutions of the SDSS Collaboration, including the Carnegie Institution for Science, Chilean National Time Allocation Committee (CNTAC) ratified researchers, Caltech, the Gotham Participation Group, Harvard University, Heidelberg University, The Flatiron Institute, The Johns Hopkins University, L'Ecole polytechnique f\'{e}d\'{e}rale de Lausanne (EPFL), Leibniz-Institut f\"{u}r Astrophysik Potsdam (AIP), Max-Planck-Institut f\"{u}r Astronomie (MPIA Heidelberg), Max-Planck-Institut f\"{u}r Extraterrestrische Physik (MPE), Nanjing University, National Astronomical Observatories of China (NAOC), New Mexico State University, The Ohio State University, Pennsylvania State University, Smithsonian Astrophysical Observatory, Space Telescope Science Institute (STScI), the Stellar Astrophysics Participation Group, Universidad Nacional Aut\'{o}noma de M\'{e}xico, University of Arizona, University of Colorado Boulder, University of Illinois at Urbana-Champaign, University of Toronto, University of Utah, University of Virginia, Yale University, and Yunnan University.

This work made use of data from ESA's Gaia mission, processed by the Gaia Data Processing and Analysis Consortium (DPAC). DPAC is funded by national institutions participating in the Gaia Multilateral Agreement.

This work made use of the Overleaf platform and the NASA Astrophysics Data System.

\section*{Data Availability}
The reddening and extinction measurements derived in this work, along with their associated uncertainties, are available for download at \url{https://doi.org/10.5281/zenodo.18239586} \citep{yu_2026_18239586}. The extinction ratios ($A_\zeta / A_V$) and coefficients ($A_{\zeta}/E(B-V)$) summarized in Table~\ref{tab:extinction} are also included in the machine-readable files provided at the same repository.

\appendix
\section{Extra Plots}
\begin{figure}
    \centering
    \includegraphics[width=0.9\columnwidth]{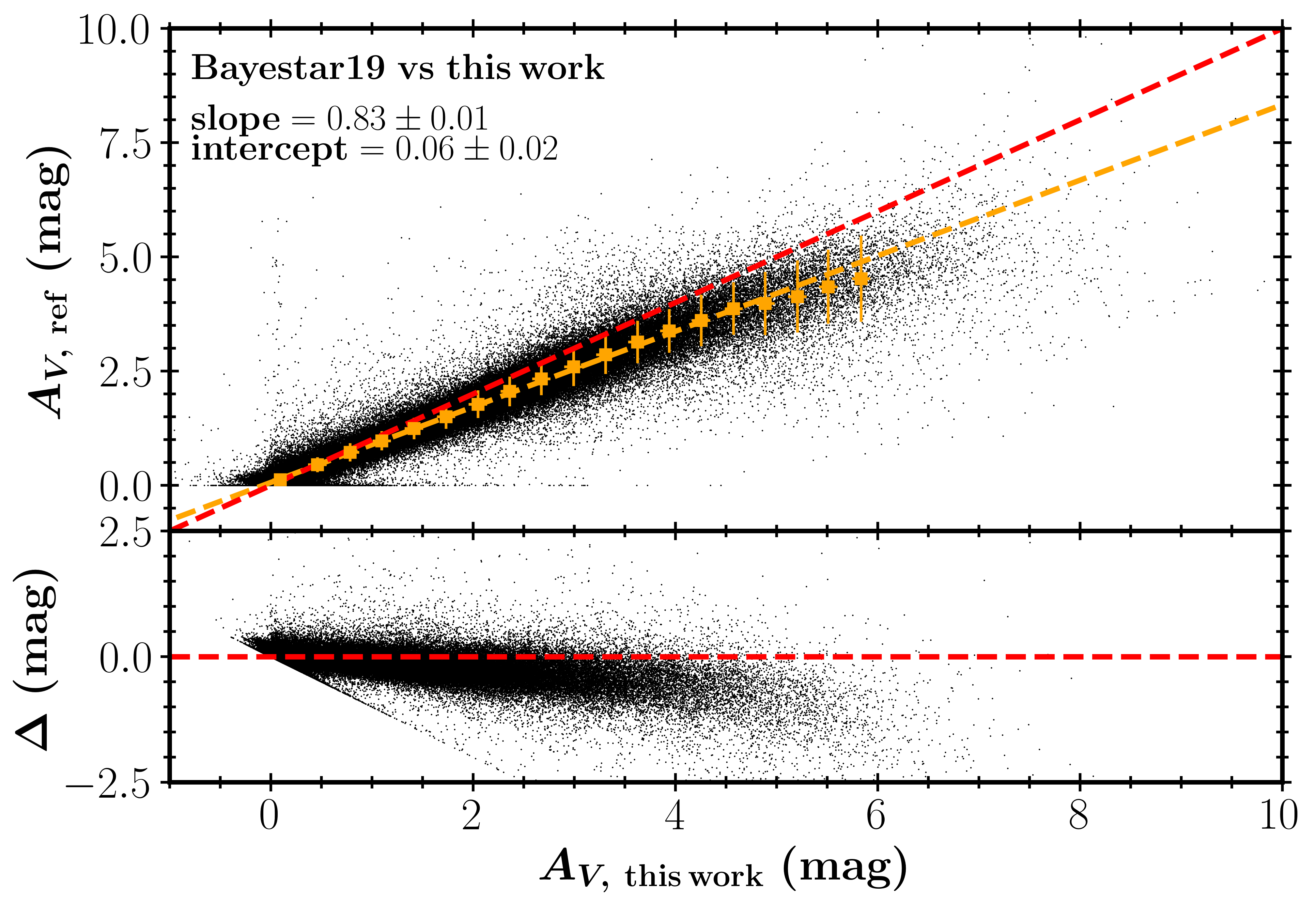}
    \caption{Comparison of $V$-band extinction values. \textbf{Upper panel:} The horizontal axis shows $A_V$ from this work, while the vertical axis shows $A_V$ from \texttt{Bayestar19}, converted from $E(B-V)$ assuming $R_V=3.1$. The red dashed line indicates the one-to-one relation. Orange squares represent the mean values in bins of width 0.32 mag; vertical error bars denote the standard deviation within each bin, while horizontal uncertainties are typically smaller than the symbol size. The orange dashed line shows the best-fitting linear relation to the binned means, accounting for their uncertainties. The slope and intercept of the fit are annotated in the panel. \textbf{Lower panel:} Residuals are defined as $\Delta = A_V(\mathrm{Bayestar19}) - A_V(\mathrm{this\ work})$. The red dashed line marks zero residual, corresponding to perfect agreement.}
\label{fig:appendix_fig1}
\end{figure}

\begin{figure}
    \centering
    \includegraphics[width=0.9\columnwidth]{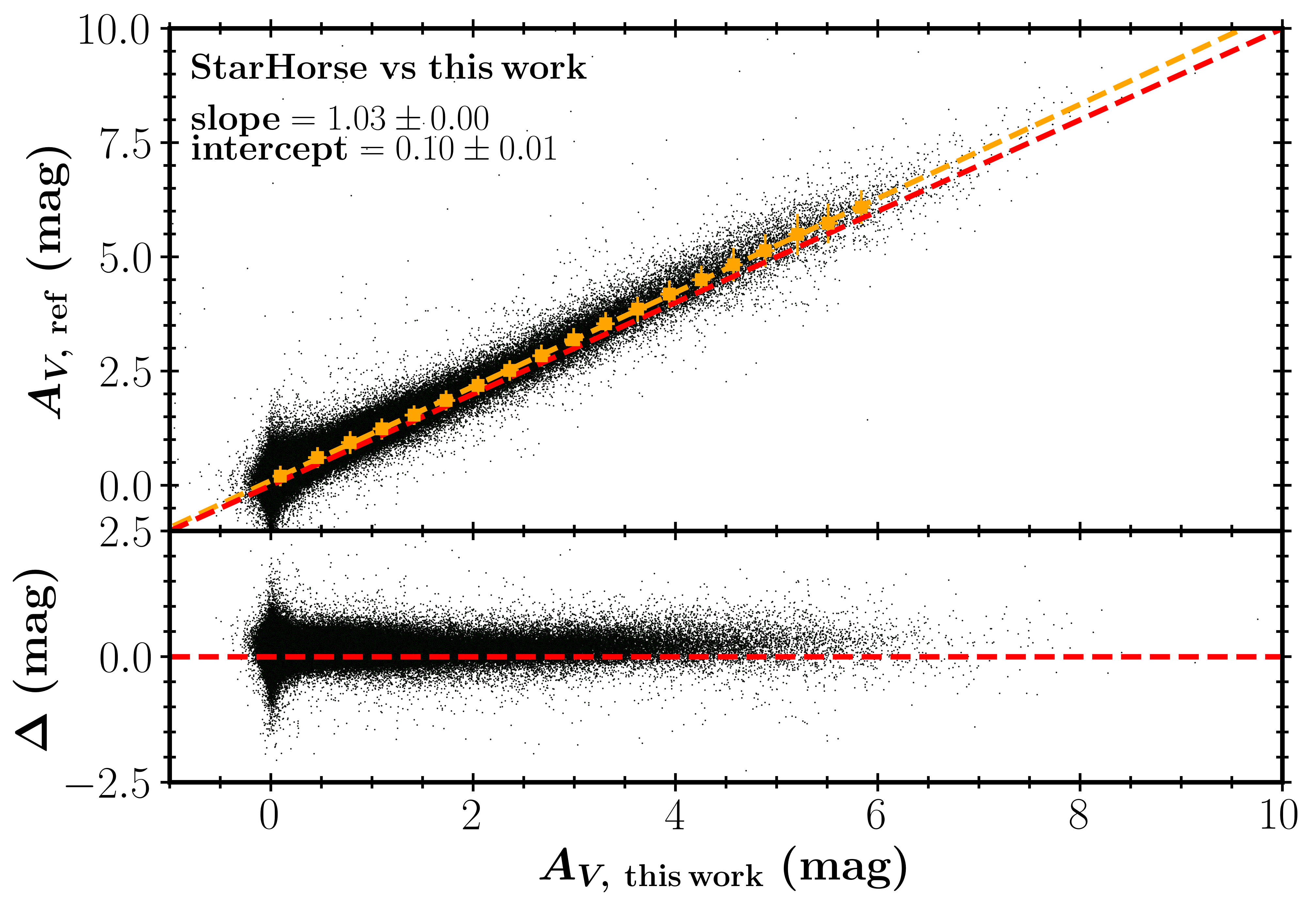}
    \caption{Same as Figure~\ref{fig:appendix_fig1}, but with reference extinction values taken from the \texttt{StarHorse} catalogue.}
    \label{fig:appendix_fig2}
\end{figure}

\begin{figure}
    \centering
    \includegraphics[width=0.9\columnwidth]{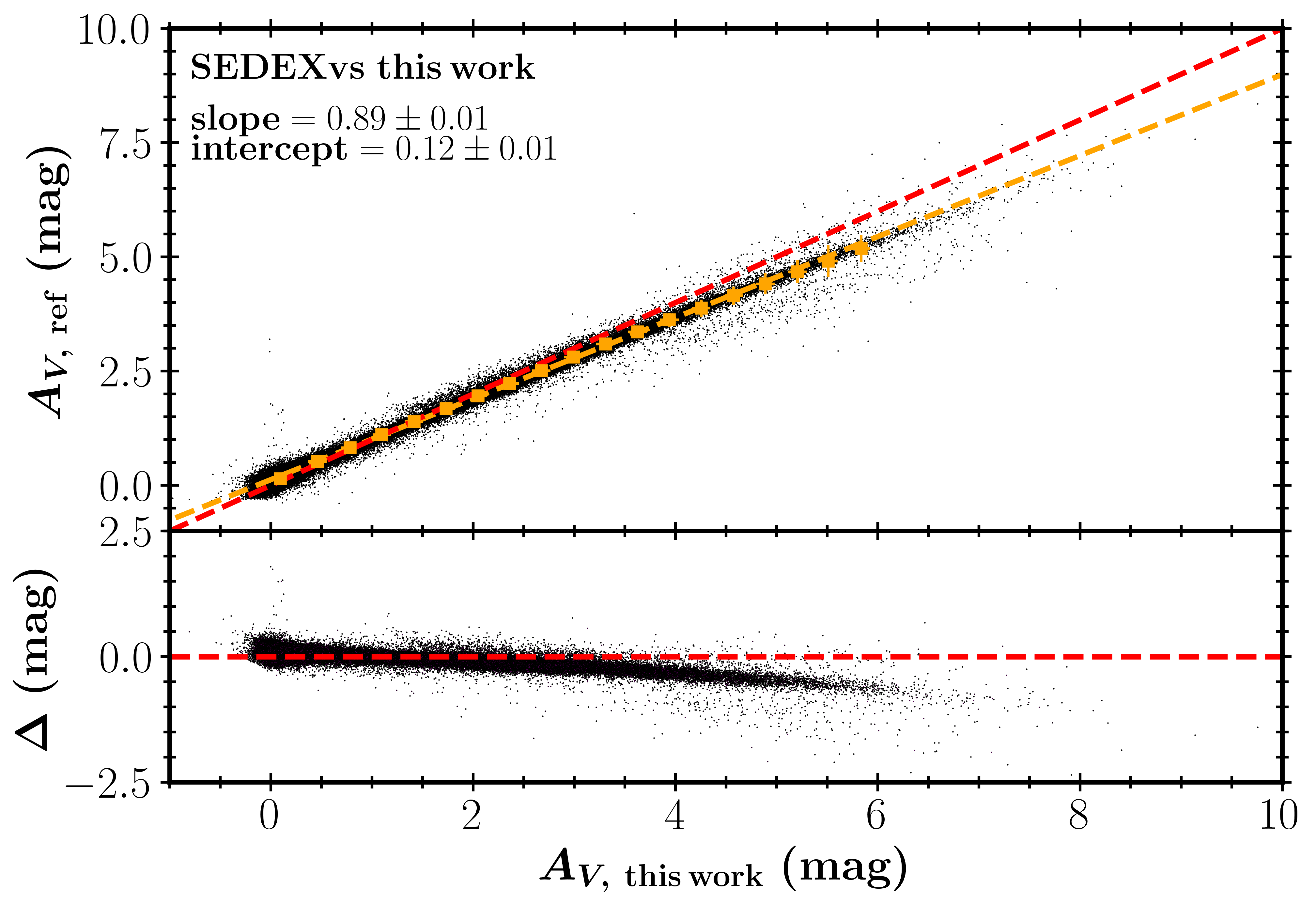}
    \caption{Same as Figure~\ref{fig:appendix_fig1}, but with reference extinction values derived using \texttt{SEDEX}.}
    \label{fig:appendix_fig3}
\end{figure}

We present comparisons between our $V$-band extinction measurements and literature results, including \texttt{Bayestar19}, \texttt{StarHorse}, and \texttt{SEDEX}. As shown in Figs.~\ref{fig:appendix_fig1}, \ref{fig:appendix_fig2}, and \ref{fig:appendix_fig3}, these comparisons provide a quantitative assessment of systematic offsets, scaling differences, and scatter between our measurements and existing extinction estimates.

\begin{figure}
    \centering
    \includegraphics[width=0.9\columnwidth]{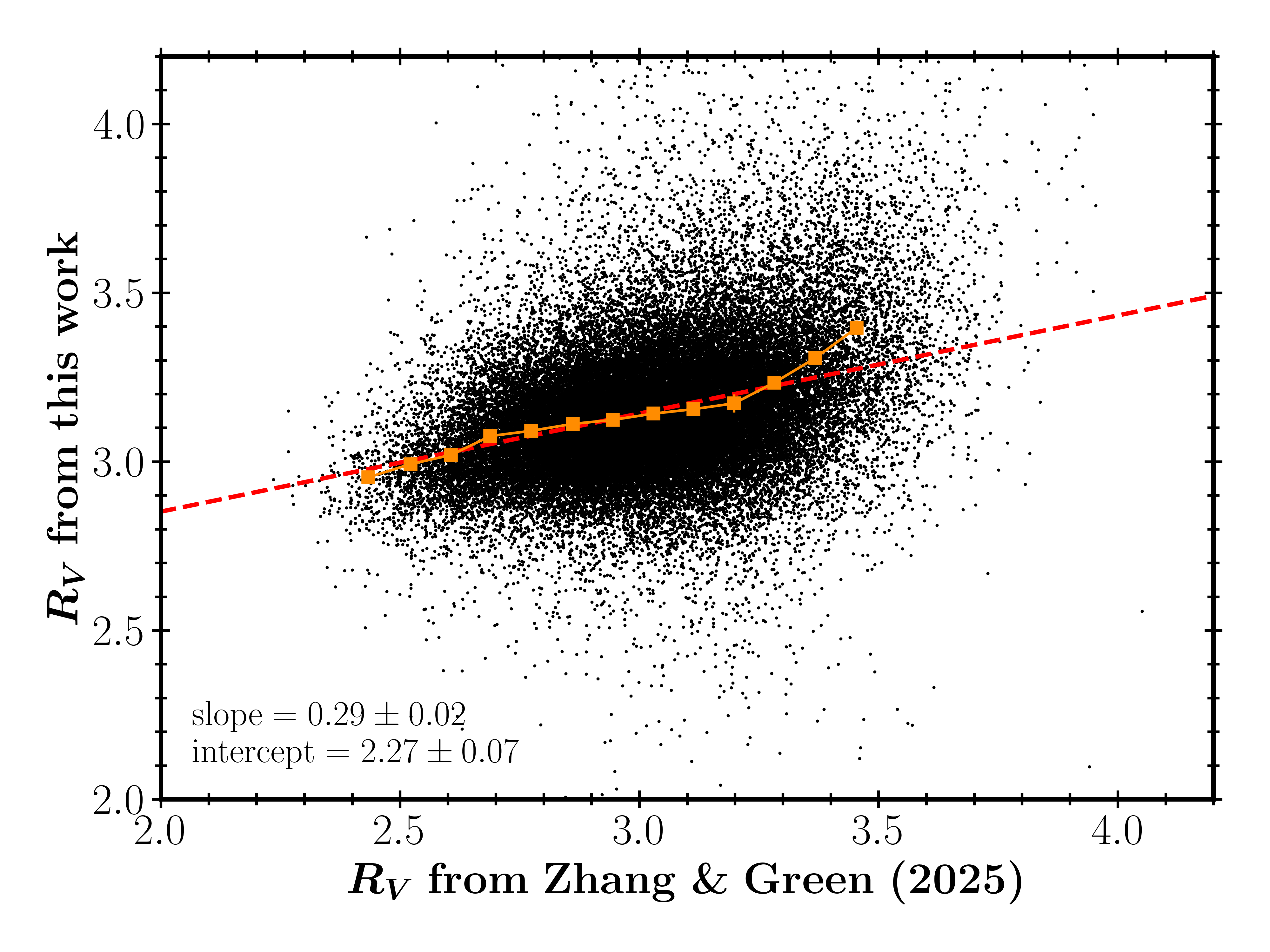}
    \caption{Comparison of our $R_V$ estimates with those predicted by the integrated 3D $R_V$ map of \citet{zhanggreen2025}, based on \textit{Gaia} BP/RP spectra. The predicted $R_V$ values are obtained by interpolating the 3D $R_V$ map at the positions of individual stars using their sky coordinates and \textit{Gaia} distances \citep{bailer-jones2021}. Orange squares indicate the mean $R_V$ values in individual bins, with standard errors smaller than the symbol size. A linear fit to the binned values is shown by the red dashed line, with the slope and intercept and their uncertainties indicated in the lower left corner.}
    \label{fig:appendix_fig4}
\end{figure}

In addition, we compare our $R_V$ measurements with those predicted by the 3D $R_V$ map of \citet{zhanggreen2025}, derived from \textit{Gaia} BP/RP spectra. The comparison is shown in Fig.~\ref{fig:appendix_fig4}.
\label{sec:appendix_plots}

\bibliographystyle{mnras}
\bibliography{references} 

\bsp 
\label{lastpage}
\end{document}